\documentclass[fleqn,twoside]{article}	
\usepackage{insa}
\usepackage{graphicx}
\usepackage{epsfig,psfig}
\newcommand{\lsim}{\raisebox{-0.13cm}{~\shortstack{$<$ \\[-0.07cm] $\sim$}}~}
\newcommand{\gsim}{\raisebox{-0.13cm}{~\shortstack{$>$ \\[-0.07cm] $\sim$}}~}

\newcommand{\ra}{\rightarrow}
\newcommand{\tb}{\tan \beta}

\newcommand{\beq}{\begin{eqnarray}}
\newcommand{\eeq}{\end{eqnarray}}

\begin{document} 

\title{Electroweak symmetry breaking at the LHC}

\author{Abdelhak Djouadi\address{Laboratoire de Physique Th\'eorique,
Universit\'e Paris--Sud,  F--91405 Orsay Cedex, France,\\
$~$ Physikalisches Institut, University of Bonn, Nussallee 12, 53115 Bonn,
Germany.} and Rohini 
Godbole\address{Center for High Energy Physics, Indian Institute of Science,
Bangalore 560 012, India.}}

\thispagestyle{empty} 
\begin{abstract} 
One of the major goals of the Large Hadron Collider is to probe the
electroweak symmetry breaking mechanism and the generation of the
masses of the elementary  particles.  We review the physics of the
Higgs sector in the Standard Model and some of its extensions such
as supersymmetric theories and models of extra dimensions.  The
prospects for  discovering the Higgs particles at the LHC and the
study of their fundamental properties are summarised.  
\end{abstract}

\maketitle

\section{Introduction}

Establishing the precise mechanism of the spontaneous breaking of
the  electroweak gauge symmetry is indeed a central focus of the
activity in the area  of high energy physics and, certainly, one of
the primary goals of the Large  Hadron Collider, the LHC, which will
soon start operation. In the Standard Model (SM), electroweak
symmetry  breaking (EWSB) is achieved via the Higgs mechanism
\cite{Higgs,Goldstone}, wherein the neutral  component of an
isodoublet scalar field acquires a non--zero vacuum  expectation
value. This gives rise to nonzero masses for the fermions and  the
electroweak gauge bosons,  which are otherwise not allowed by the 
${\rm SU(2)\!\times\! U(1)}$ symmetry. In the sector of the theory
with broken symmetry, one of the four degrees of freedom of the
original isodoublet field, corresponds to a physical particle:  the
Higgs  boson with  ${\rm J^{PC}}\!=\!0^{++}$ quantum numbers under
parity and charge conjugation   \cite{HHG,Djouadi:2005gi}.

Clearly, the discovery of this last missing piece of the SM is a
matter of profound importance.  In fact, in spite of the phenomenal
success of the SM  in explaining the precision data~\cite{LEPEWWG}, 
the SM can not be considered to be established  completely until the
Higgs particle is  observed experimentally and, further, its
fundamental properties such as  its mass, spin and other quantum
numbers, as well as its couplings to  various matter and gauge
particles and its self-couplings are established. These studies are 
important not only to establish the SM as the correct theory of
fundamental particles and interactions among them, but also to 
achieve further clarity into the dynamics of the EWSB mechanism. 

Indeed, in spite of the success of the idea of spontaneous symmetry
breaking (in fact, partly honoured by the Nobel prize of 2008) in
different areas of physics, very little is known about its
realisation in particle physics via the Higgs mechanism. The many
important questions which one would like answered are: does the
dynamics involve new strong interactions and is the Higgs a
composite field? if elementary Higgs particles indeed exist  in
nature, how many fields are there and in which gauge representations
do  they appear? does the EWSB sector involve sizable CP violation?
etc. 

Theoretical realizations span a wide range of scenarios extending
from weak  to strong breaking mechanisms, including the so called
Higgsless theories in extra dimensional models. As far as the
representations of the gauge group  are concerned, there is again a
whole range starting from models involving  light fundamental Higgs
fields, arising from an SU(2) doublet,  such as in the  SM and its
supersymmetric extensions which include two--Higgs doublets in the
minimal version, to those containing additional singlet fields or
higher  representations in extended versions in unified theories
and/or alternative theories such as little Higgs models. 

Furthermore, the link between particle physics and cosmology means
that the EWSB mechanism can have implications for the  generation
of  the baryon--antibaryon asymmetry in the early universe and could
play an important role in the annihilation of the new particles that
are responsible for the cosmological dark matter and thus impact
their density in the universe today. In fact,  possible CP
violation  in the Higgs sector can have a direct bearing on the two
cosmology issues mentioned above.  An understanding of the EWSB
mechanism at a more fundamental level might also hold clues about
why the three generations of quarks and leptons have masses which
differ from each other; the so called flavour issue. 

A complete discussion of Higgs physics thus touches upon almost all
the issues under active investigation in theoretical  and
experimental particle physics.

\section{Electroweak symmetry breaking mechanism}

\subsection{The Higgs boson in the SM}\smallskip

In the SM there exists only one isodoublet complex scalar field and,
thus, there are initially four real scalar fields
\cite{Higgs,Goldstone,HHG,Djouadi:2005gi}. After spontaneous EWSB,
we are left with one physical degree of freedom, the Higgs scalar
and the other three  would--be Nambu-Goldstone bosons are absorbed
to build up the longitudinal components of the $W^\pm,Z$ gauge
bosons and generate their masses. Yukawa interactions of the
fermions  with the same scalar field give rise to the fermion
masses. The Higgs scalar has ${\rm J^{\rm PC}=0^{++}}$  assignments
of spin, parity and charge conjugation quantum  numbers. The Higgs
couplings to the fermions and gauge bosons are  related to the 
masses of these particles and are thus decided by the symmetry
breaking mechanism. In contrast, the mass of the Higgs boson itself
is completely undetermined in the model. There are, however, both
experimental and theoretical constraints on this fundamental
parameter, which we will summarize below.

One available direct information on the Higgs mass is the lower
limit  $M_H \gsim 114.4$ GeV at 95\% confidence level (c.l.)
established at LEP2 \cite{Barate:2003sz}. The collaborations have
also reported a small, $\lsim 2 \sigma$, excess of events beyond the
expected SM backgrounds, consistent with a SM--like Higgs boson with
a mass $M_H \sim 115$ GeV \cite{Barate:2003sz}. In addition to this,
the Tevatron physics potential for the discovery of Higgs particles
looks promising, with the coming larger data sets. In particular,
evidence for the SM Higgs boson could be obtained if the mass is
near  the observed experimental lower limit from LEP of about $115$
GeV or if it is near 160 GeV. In fact, with the run-II data
collected by both the experiments, corresponding to $2.5$ fb$^{-1}$,
the observed upper  limits are a factor $3.7\, (1.1)$ higher than
the expected SM Higgs cross section at $M_H = 115\, (160)$ GeV at
$95 \%$ c.l.~\cite{tevreview}.

Furthermore, the high accuracy of the electroweak data measured at
LEP, SLC and the Tevatron \cite{PDG}  provides an indirect
sensitivity to $M_H$:  the Higgs boson contributes logarithmically,
$\propto \log (M_H/M_W)$, to the radiative corrections to the $W/Z$
boson propagators. A recent analysis, which uses the updated 
determination  of the top quark mass ($172.4$ GeV), yields the value
$M_H=84^{+34}_{-26}$ GeV, corresponding to a 95\% confidence level
upper limit of $M_H \lsim 154$ GeV \cite{LEPEWWG}. A very recent
analysis, using a new fitting  program gives the more precise value
$M_H=116.4 ^{+18.3}_{-1.3}$ GeV \cite{Gfitter}.

\begin{figure}[!h]
\begin{center}
\vspace{-0.5cm}
\hspace*{-1cm}
\includegraphics[width=7cm,height=7cm] {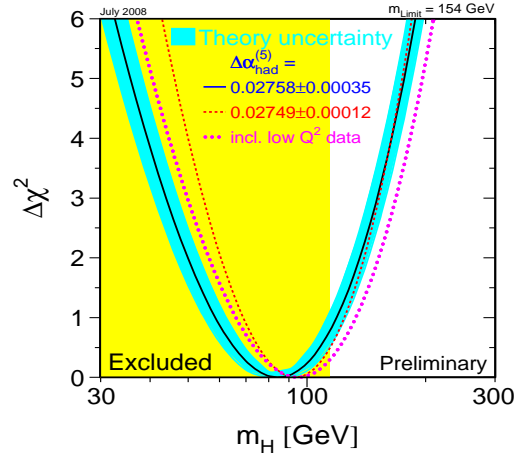}
\vspace{-1.5cm}
\end{center}
\caption{Experimental limits on the mass of the Higgs boson in the 
SM from a global fit to the electroweak precision data; the excluded
region from direct Higgs searches is also shown~\cite{LEPEWWG}.}
\label{Hmass:expconstr}
\vspace{-0.5cm}
\end{figure}

Fig.~\ref{Hmass:expconstr} shows  the  global fit to the
electroweak data and  the ensuing limit on the Higgs mass. The
limit increases to $185$ GeV when the LEP2 direct search limit of
$114$ GeV is included. If the Higgs boson turns out to be
significantly heavier than these upper limits, $M_H \gsim 200$ 
GeV,   there should be an additional new  ingredient that is
relevant at the EWSB scale which  the next round of  experiments
should reveal.

From the theoretical side, interesting constraints can be derived
from assumptions on the energy range within which the SM is valid
before  perturbation theory breaks down and new phenomena would
emerge.   For instance, if the Higgs mass were larger than $\sim$ 1
TeV, the $W$ and $Z$ bosons would have to interact very strongly
with each other so that their scattering at high energies respects 
unitarity. Imposing the unitarity requirement in the high--energy
scattering of gauge bosons leads to the  bound $M_H \lsim 700$ GeV
\cite{H-LQT}.  If the  Higgs boson were too heavy, unitarity would
be violated in these processes at  energies above $\sqrt s \gsim 
1.2$ TeV and new phenomena should appear to  restore it.  It is interesting 
to note, as an aside, that just the requirement of perturbative 
unitarity in $WW$ scattering leads to a model with exactly 
the same particle content and couplings as the SM~\cite{cornwall}.

Another important theoretical constraint comes from the fact that
the quartic Higgs self--coupling, which at the scale $M_H$ is fixed
by $M_H$ itself, grows logarithmically with the energy scale. If
$M_H$ is small, the energy cut--off $\Lambda$ at which the coupling
grows beyond any bound and new phenomena should occur, is large; if
$M_H$ is large, the cut--off $\Lambda$ is small.  The condition $M_H
\lsim \Lambda$ sets an upper limit on the Higgs mass in the SM, the
so called triviality bound. A naive one--loop analysis assuming the 
validity of perturbation theory \cite{Roman-plot} as well as lattice
simulations \cite{H-Lattice} lead to the estimate $M_H \lsim 630$
GeV for this limit.  Furthermore, loops involving top quarks tend to
drive the coupling to negative values for which the vacuum  is no
longer stable.

\begin{figure}[t]
\begin{center}
\vspace{-.5cm}
\includegraphics*[width=9cm,height=6cm] {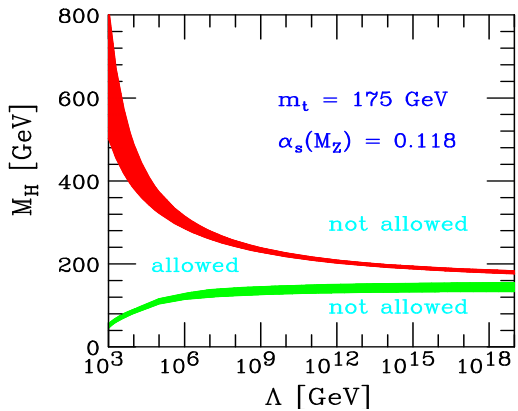}
\vspace{-1.5cm}
\end{center}
\caption{Theoretical upper and lower bounds on the Higgs mass in 
the SM from the assumption that the SM is valid up to the cut--off
scale $\Lambda$ \cite{Hambye:1996wb}.}
\label{Hmass:thconstr}
\vspace{-0.8cm}
\end{figure}

Requiring the SM to be extended to, for instance, the GUT scale
$\Lambda_{\rm GUT} \sim 10^{16}$ GeV and including the effect of top
quark loops on the running coupling, the Higgs boson mass should lie
in the range 130 GeV $\lsim M_H \lsim 180$ GeV \cite{Hambye:1996wb};
see  Fig.~\ref{Hmass:thconstr}. In fact, in any model beyond the SM
in which the theory is required to be weakly interacting up to the
GUT or Planck scales, the Higgs boson should be lighter than $M_H
\lsim 200$ GeV. Such a Higgs particle can be thus produced at the
LHC. 


Once its mass is fixed the profile of the Higgs particle is uniquely
determined and its production rates and decay widths are fixed.  As
its couplings to different particles are proportional to their masses, 
the Higgs boson  will have the tendency to decay into the heaviest particles
allowed by phase space. The Higgs decay modes and their branching
ratios (BR) are briefly summarized below; see Ref.~\cite{decays}
for details. 

In the ``low--mass" range, $M_H \lsim 130$ GeV,  the Higgs boson
decays into  a large variety of channels. The main mode is by far the
decay into  $b\bar{b}$ with BR\,$\sim$ 90\% followed by the decays
into $c\bar{c}$ and  $\tau^+\tau^-$ with BRs\,$\sim$ 5\%. Also of
significance is the top--loop  mediated decay into gluons, which
occurs at the level of $\sim$ 5\%.  The top and $W$--loop mediated
$\gamma\gamma$ and $Z \gamma$ decay modes, which lead to  clear
signals, are  very rare with BRs of ${\cal O}(10^{-3})$.  

In the ``high--mass" range, $M_H \gsim 130$ GeV, the Higgs bosons
decay into $WW$ and $ZZ$ pairs, one of the gauge bosons being possibly
virtual  below the thresholds. Above the $ZZ$ threshold, the BRs are
2/3 for $WW$ and  1/3 for $ZZ$ decays, and the opening of the
$t\bar{t}$ channel for higher $M_H$ does not alter  this pattern
significantly.  

In the low--mass range, the Higgs is very narrow, with $\Gamma_H<10$
MeV, but this width increases, reaching 1 GeV at the $ZZ$ threshold.
For  very large masses, the Higgs  becomes obese, since $\Gamma_H \sim
M_H$, and can hardly be considered as a resonance. 

\begin{figure}[!h]
\vspace{-1cm}
\begin{center}
\includegraphics*[width=7cm,height=5cm]{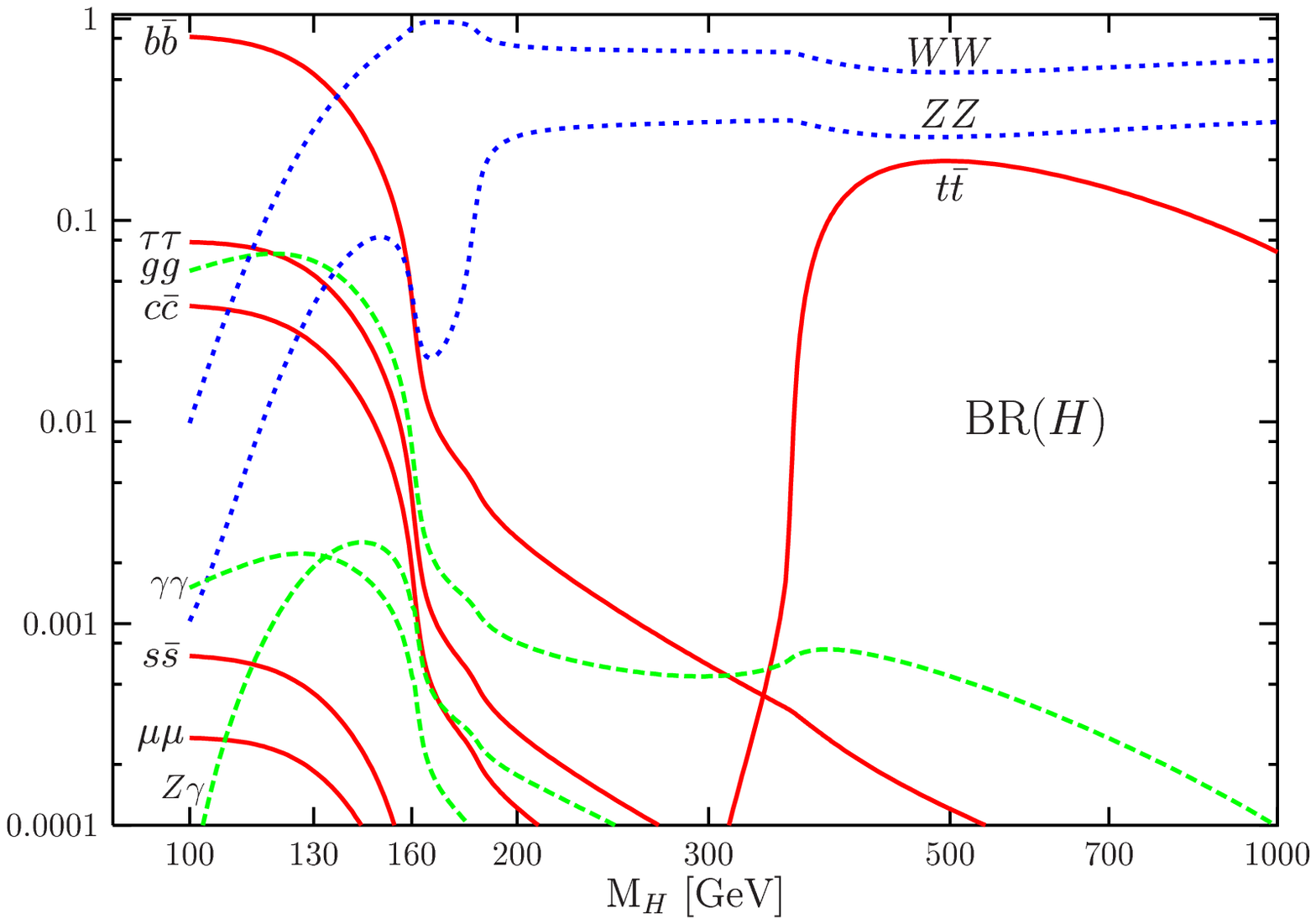}\\
\includegraphics*[width=7cm,height=5cm]{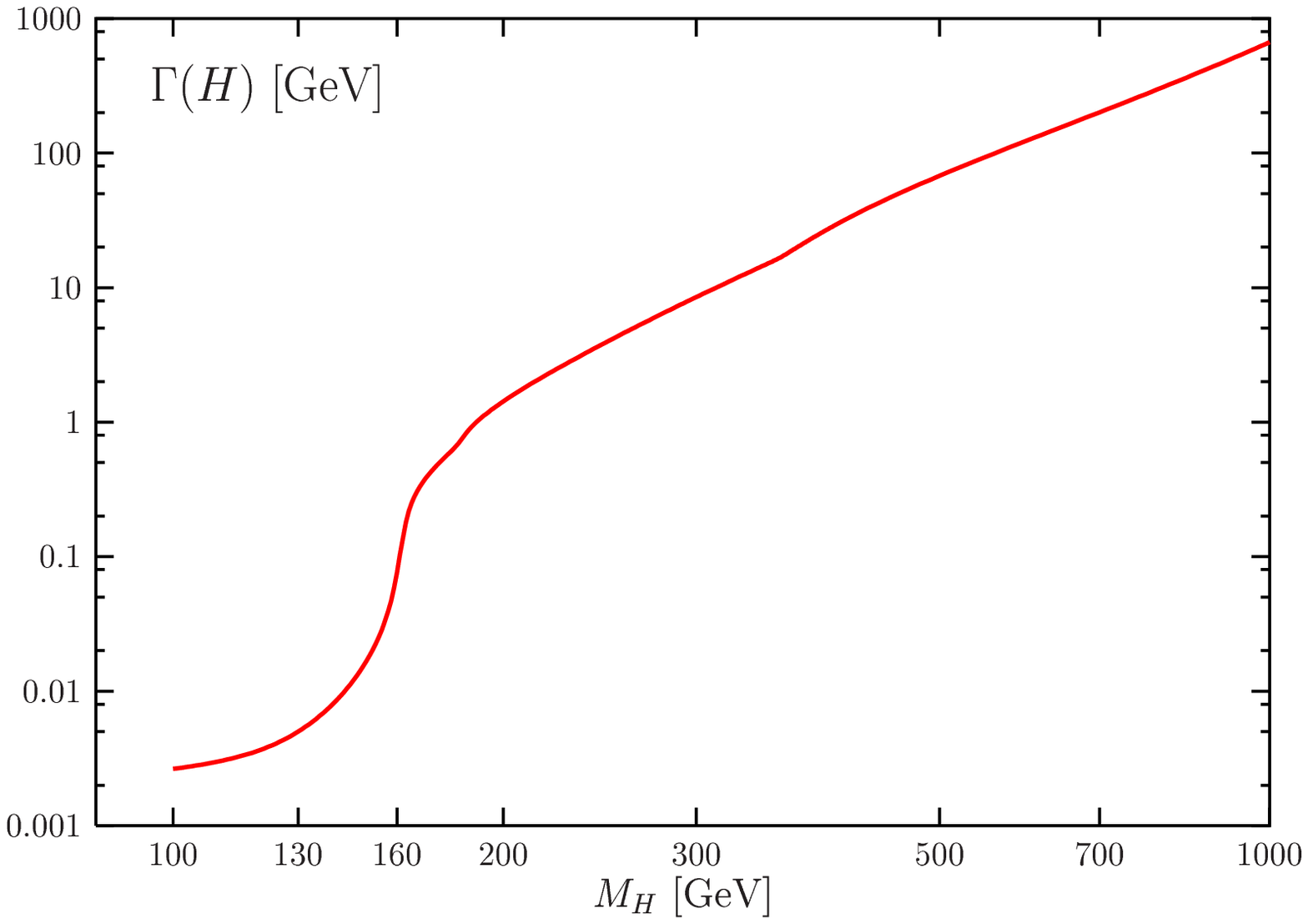}
\end{center}
\vspace{-1.cm}
\caption{The decay branching ratios (top) and the total decay width (bottom) 
of the SM Higgs boson as a function of its mass \cite{hdecay}.} 
\label{Hfig:brwidsm}
\vspace*{-.7cm}
\end{figure}

The branching ratios and total decay widths are summarized in
Fig.~\ref{Hfig:brwidsm}, which is  obtained from a  recently updated
version of the code {\tt HDECAY} \cite{hdecay} and where the  new
value $m_t=172$ GeV is used as an input. 

The SM in spite of its spectacular success, is ridden with two well
known problems, which are the major stumbling blocks while trying
to extend  the validity of the SM to the GUT scale $\Lambda_{\rm
GUT}$. The first one  is the so--called  naturalness problem:  the
radiative corrections to $M_H$ being quadratically divergent  push
the Higgs boson  mass to be  the order of these large scales.  The
second problem is that the running of the three gauge couplings of
the  SM is such that they do not meet at a single point and thus do
not unify at the GUT scale. 

Low energy supersymmetry solves these two problems at once
\cite{MSSMbook}: supersymmetric particle loops cancel exactly the
quadratic divergences and help stablise the Higgs boson mass at the
weak scale, and they contribute to the running of the gauge
couplings to allow their unification at $\Lambda_{\rm GUT}$. In
addition, it allows for a good candidate for the dark matter in the
universe.

\subsection{The Higgs particles in the MSSM}\smallskip

The minimal supersymmetric extension of the SM (MSSM), requires the
existence of two isodoublet  Higgs fields to cancel anomalies and to
give mass separately to up and down--type fermions.  Two CP--even
neutral Higgs bosons $h,H$, a pseudoscalar $A$ boson  and a pair of
charged scalar particles, $H^\pm$, are introduced by this  extension
of the Higgs sector \cite{HHG,MSSMbook,Djouadi:2005gj}. In fact, in
this case, the scalar  potential  does not involve an arbitrary self
coupling  $\lambda$ as is the case with the SM, but involves only
the gauge couplings and as a result  the mass of the lightest Higgs
boson $h$ is bounded from above. Besides the  four masses, the
properties of the Higgs sector in the MSSM are determined by  two
more parameters  : a mixing angle $\alpha$ in the neutral CP--even
sector  and the ratio of the two vacuum expectation values $\tb$. 
The value of the  latter lies in the range $1 \lsim \tb \lsim
m_t/m_b$.

\begin{figure}[!h]
\begin{center}
\vspace*{-8.mm}
\hspace*{-4mm}
\includegraphics[width=4.55cm]{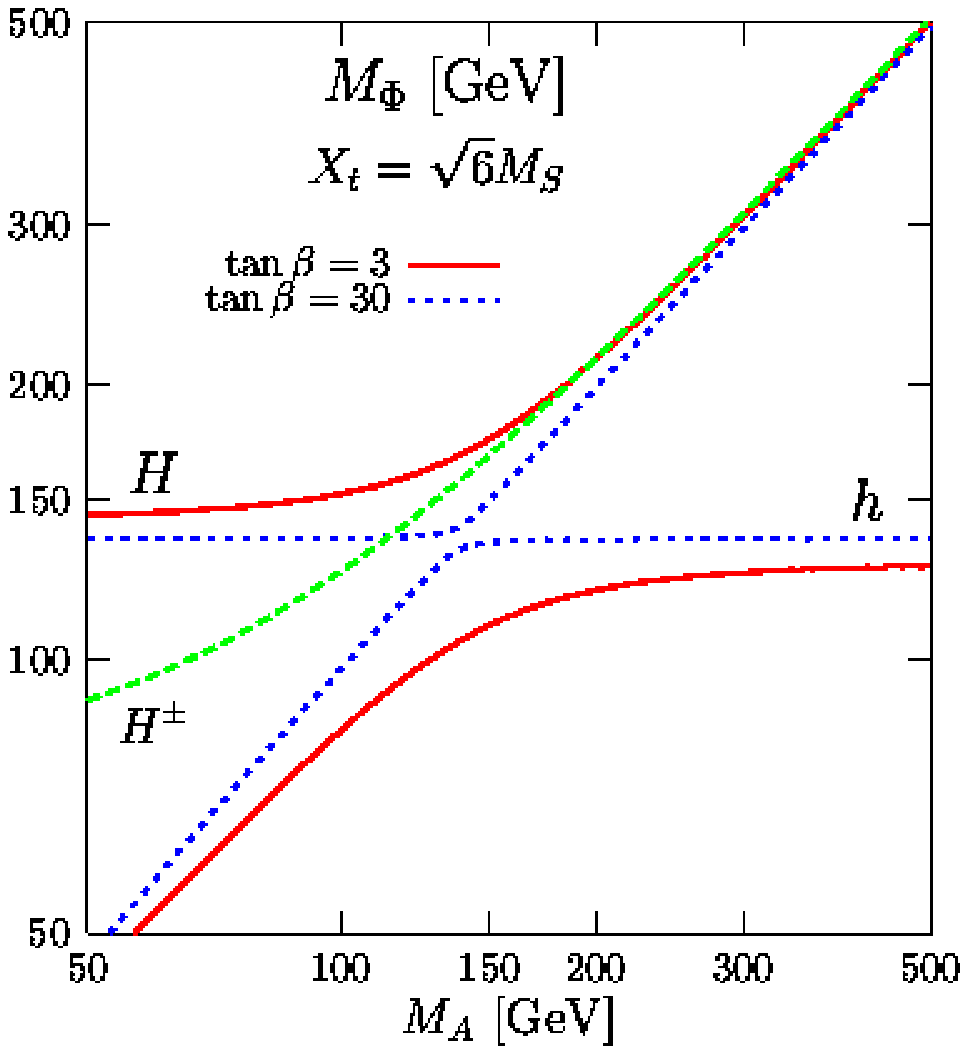}\hspace{-0.7cm}
\includegraphics[width=4.55cm]{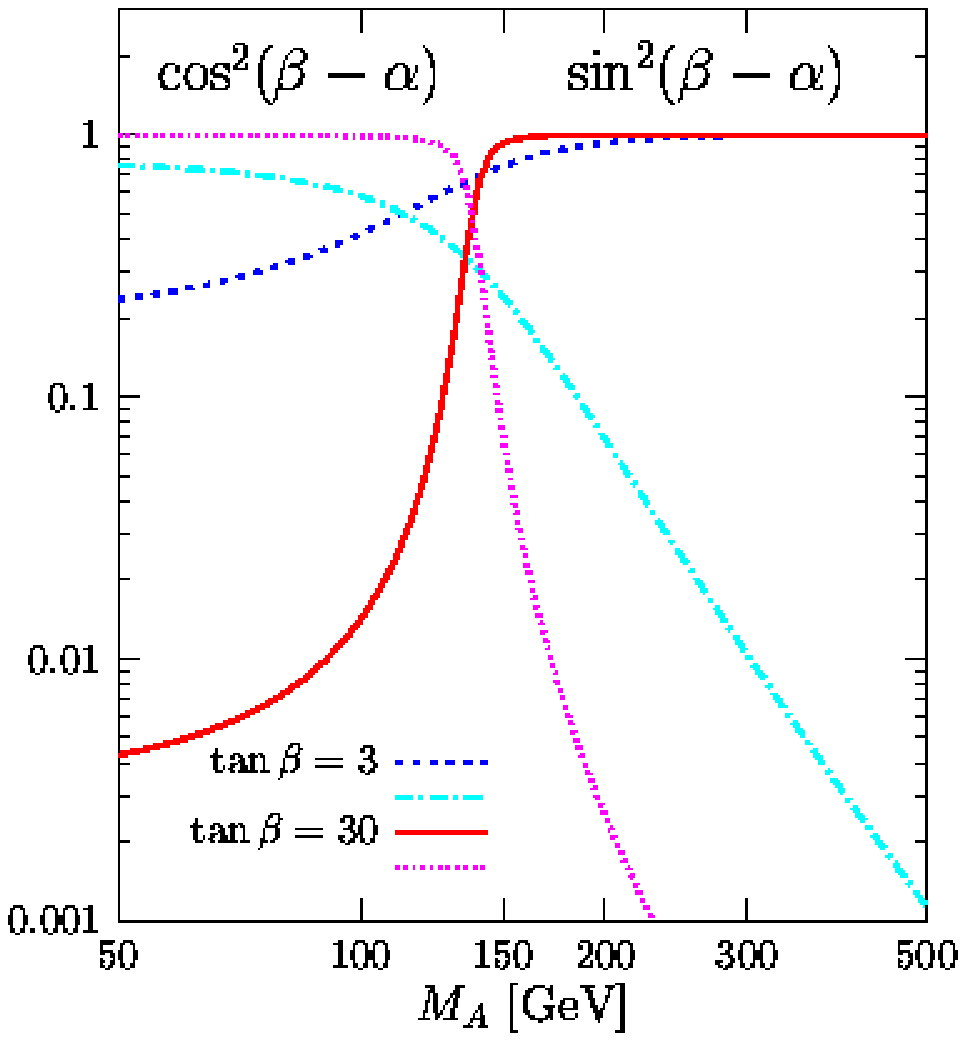}\hspace{-0.8cm}
\end{center}
\vspace*{-1.5cm}
\caption{The masses and coupling of the Higgs bosons in the MSSM:
as a function of $M_A$ for $\tb=3,30$ with $M_S=2$ TeV and
$X_t=\sqrt 6 M_S$.}
\label{Hfig:mssmmass_coup}
\vspace*{-1.5cm}
\end{figure}

Supersymmetry leads to several relations among these parameters and
only two of them, taken in general to be $M_A$ and $\tb$, are in
fact independent. These relations impose a strong hierarchical
structure on the mass spectrum, $M_h\!<\!M_Z, M_A\!<\!M_H$ and
$M_W\!<\!M_{H^\pm}$, which however is broken by radiative
corrections as the top quark mass is large; see
Ref.~\cite{gigazsven} for a review. The leading part of this
correction grows as the fourth power of $m_t$ and logarithmically
with the SUSY scale  or common squark mass $M_S$; the mixing (or
trilinear coupling) in the stop sector $A_t$ plays an important
role. For instance, the upper bound on the mass of the lightest
Higgs boson $h$ is shifted from the tree level value $M_Z$ to $M_h
\sim 130$--140 GeV in the maximal mixing scenario where $X_t= A_t
-\mu/\tb \sim 2M_S$ with $M_S={\cal O}(1$ TeV) \cite{gigazsven};
see  left panel of  Fig.~\ref{Hfig:mssmmass_coup}. The masses of the
heavy neutral and charged Higgs particles are expected to range from
$M_Z$ to the SUSY breaking scale $M_S$.

The pseudoscalar Higgs boson $A$  has no tree level couplings to
gauge bosons, and its couplings to down (up) type fermions are
(inversely) proportional to $\tb$. This is also the case for the
couplings of the charged Higgs boson to fermions, which are
admixtures of scalar and pseudoscalar currents and depend only on
$\tb$. For the CP--even Higgs bosons $h$ and $H$, the couplings to
down (up) type fermions are enhanced (suppressed) compared to the SM
Higgs couplings for $\tb >1$. They share the SM Higgs couplings to
vector bosons as they are suppressed by $\sin$ and
$\cos(\beta-\alpha)$ factors, respectively for $h$ and $H$.  The
Higgs couplings to the $W^\pm, Z$ bosons are  displayed in the right
panel of Fig.~\ref{Hfig:mssmmass_coup}.

If the pseudoscalar mass is large, the $h$ boson mass reaches its
upper limit [which, depending on the value of $\tb$ and stop mixing,
is in the range 100--140 GeV] and its couplings to fermions and
gauge bosons are SM--like; the heavier CP--even $H$ and charged
$H^\pm$ bosons become degenerate with the pseudoscalar $A$ boson and
have couplings to fermions and gauge bosons of the same intensity.
In this \underline{decoupling limit}, which can be already reached
for  masses $M_A \gsim 300$ GeV, it is very difficult
to distinguish the Higgs sectors of the SM and MSSM if only the
lighter $h$ particle is observed.  

\begin{figure}[!h]
\begin{center}
\vspace*{-8mm}
\includegraphics*[width=7cm,height=5cm]{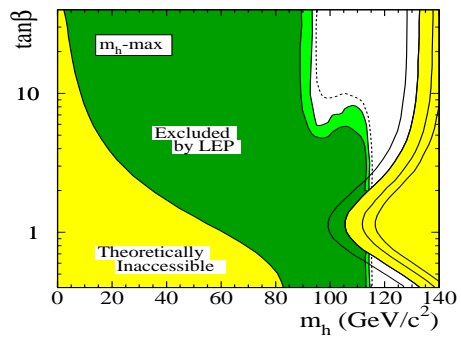}
\end{center}
\vspace*{-11mm}
\caption{The MSSM exclusion contours from LEP at $95\%$ (light
green) and $99.7\%$c.l. (dark green) for the $m_h$--max 
scenario in the $m_h$--$\tan \beta$ plane \cite{LEP2-HMSSM}.}
\label{Hfig:mssmleplim}
\vspace*{-8mm}
\end{figure}

Finally we note the experimental constraints on the
MSSM Higgs masses,  coming  mainly from the negative LEP2 searches
\cite{LEP2-HMSSM}. In the decoupling limit where the $h$ boson is
SM--like, the limit $M_h \gsim 114$ GeV from the Higgs--strahlung
$e^+ e^- \to hZ$ process holds; this constraint rules out $\tb$
values smaller than $\tb \sim 3$. Combining all processes,  the
current limits in the CP conserving MSSM at $95 \%$ c.l., assuming
no invisible decays, are~\cite{PDG,LEP2-HMSSM}: $M_h > 92.4$ GeV,
$M_A > 93.4$ GeV for $\tan \beta > 0.4$ and  $M_{H^\pm} > 79.3$ GeV.
Fig.~\ref{Hfig:mssmleplim}  shows the current limits from LEP and
Tevatron data on the MSSM Higgs sector.


\begin{figure}[!h]
\begin{center}
\includegraphics[width=8cm,height=5cm]{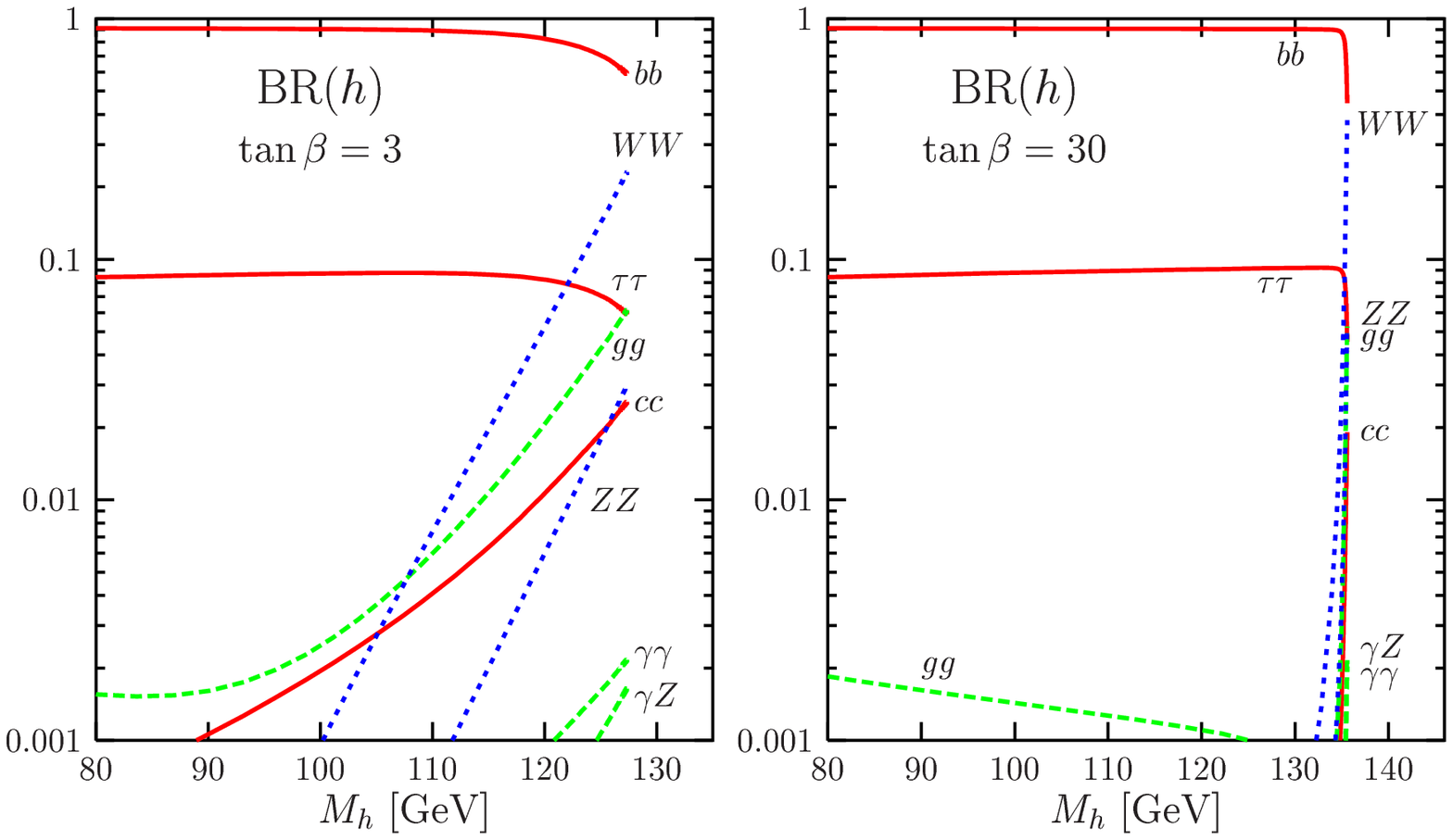}\vspace{-.3cm}
\includegraphics[width=8cm,height=5cm]{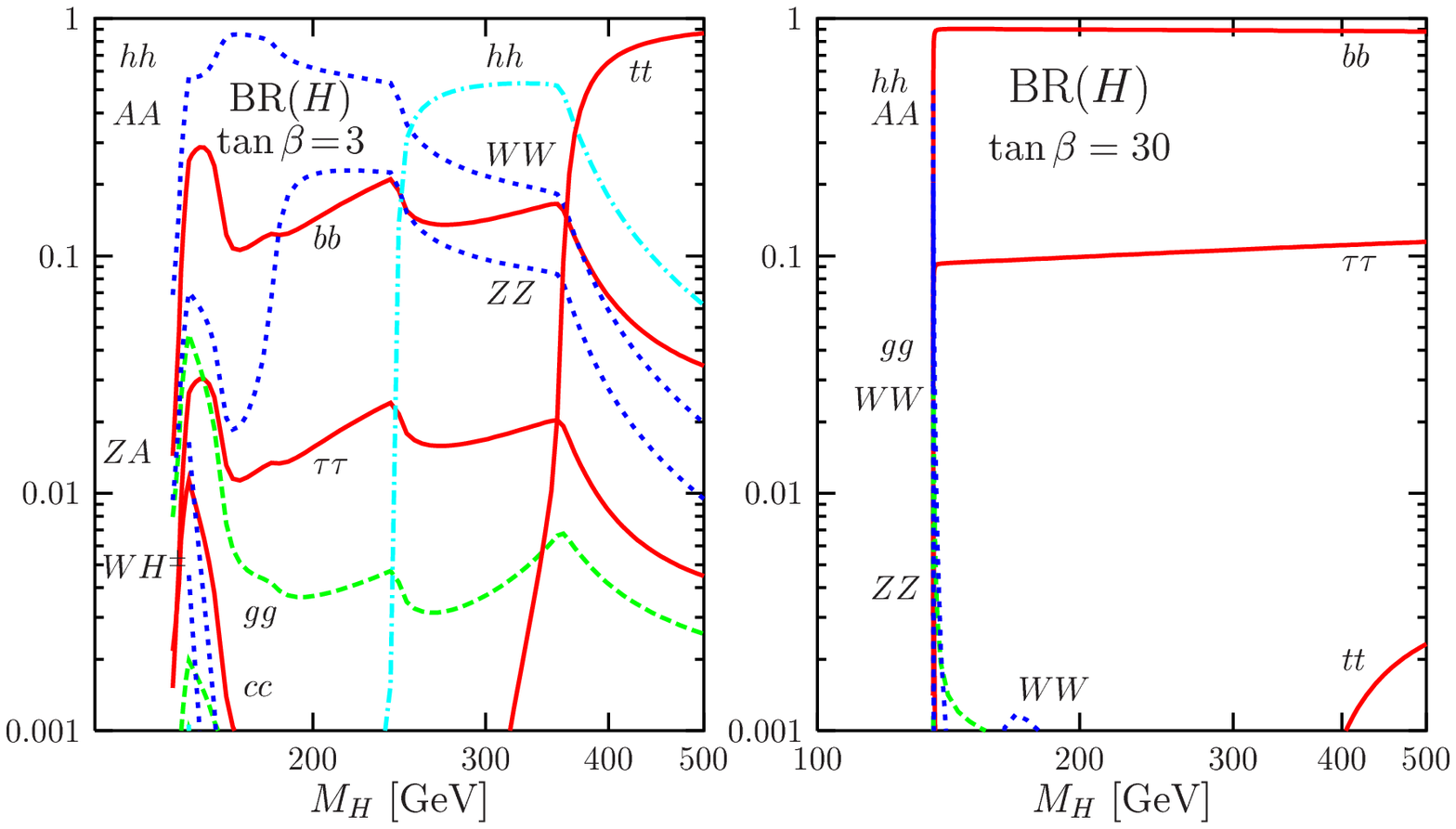}\vspace{-.3cm}
\includegraphics[width =8cm,height=5cm]{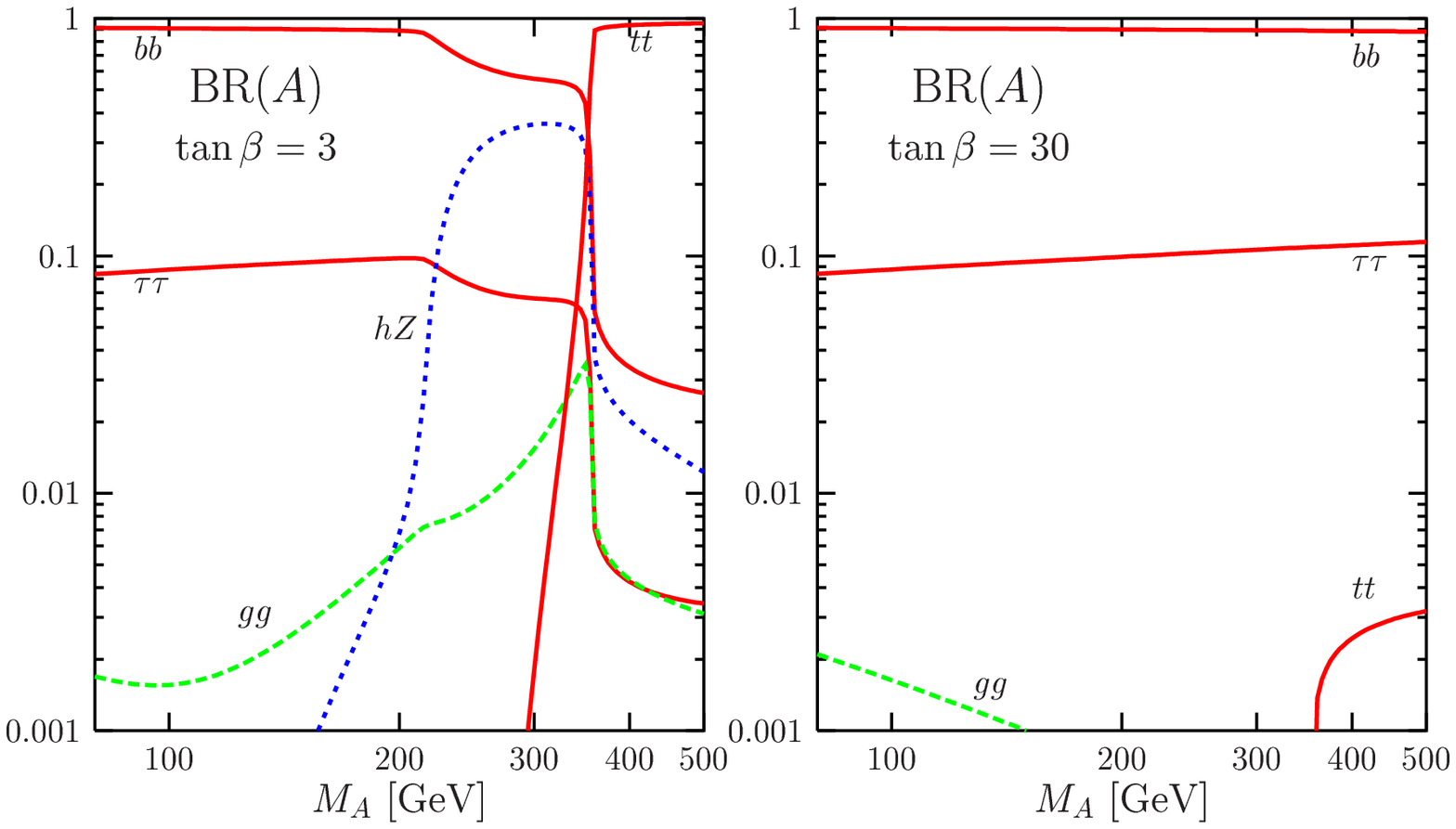}\vspace{-.3cm} 
\includegraphics[width=8cm,height=5cm]{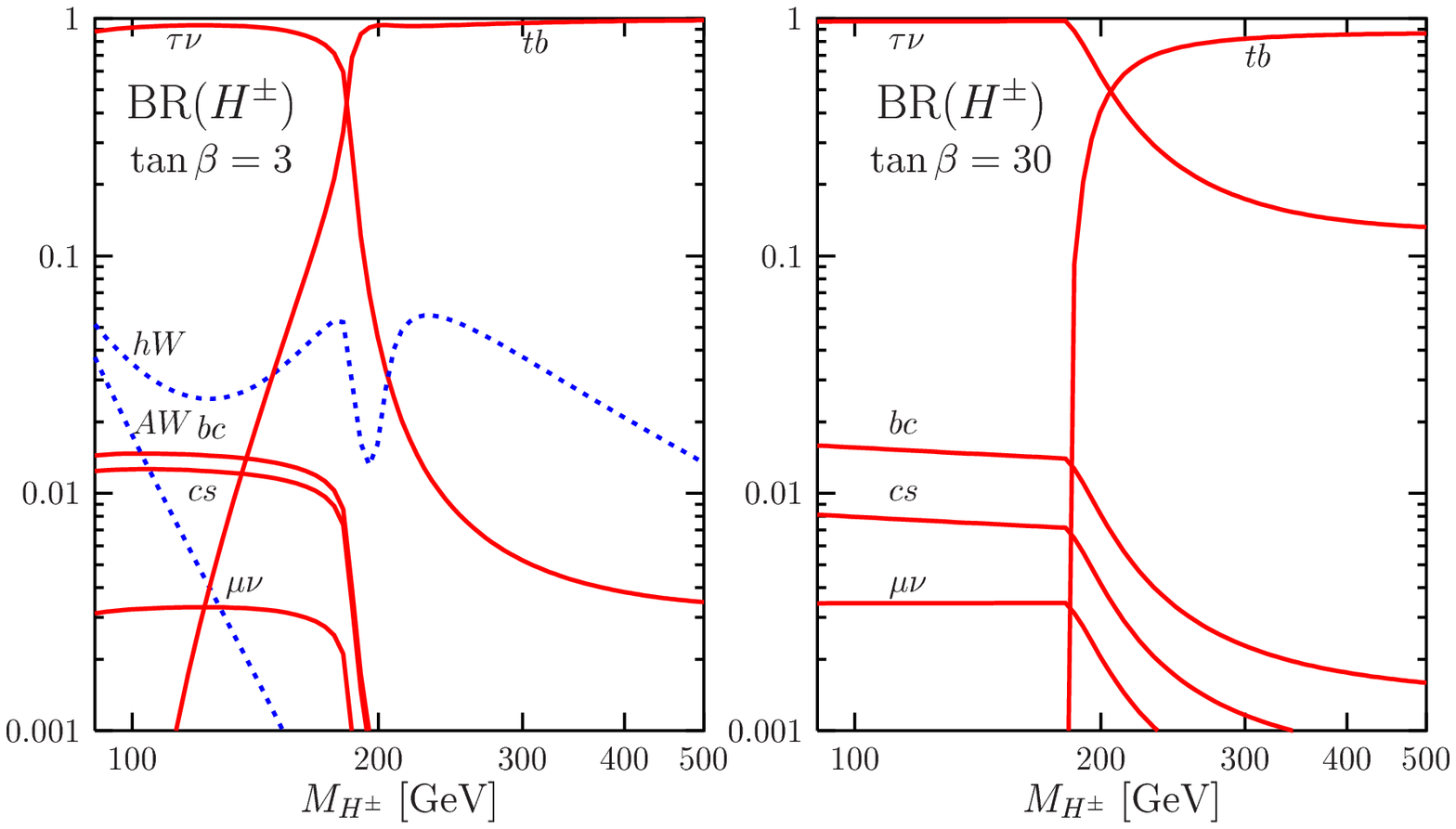}\vspace*{-1.2cm}
\end{center}
\caption{The decay branching ratios  of the  MSSM Higgs bosons as 
functions of their masses for $\tb=3$ and 30 as obtained with an 
update of  {\tt HDECAY} \cite{hdecay}; $m_t=172$ GeV and the maximal 
mixing scenario $X_t=\sqrt{6}M_S$ with $M_S=2$ TeV are assumed.} 
\label{Hdecay:MSSM}
\vspace{-.7cm}
\end{figure}

\begin{figure}[!h]
\begin{center}
\vspace*{-1.cm}
\includegraphics[width=8cm,height=5cm]{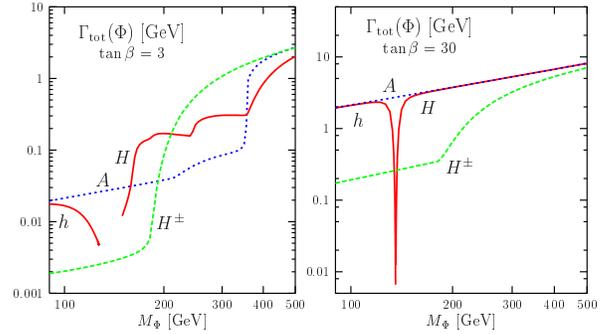}
\vspace*{-1.2cm}
\end{center}
\caption{The  total widths of the MSSM Higgs bosons as functions of
their masses for the inputs of Fig.~\ref{Hdecay:MSSM}.} 
\label{Hdecay:MSSMw}
\vspace{-.7cm}
\end{figure}

Let us now turn to the decays of the MSSM Higgs particles;
Fig.~\ref{Hdecay:MSSM}. The lightest $h$ boson will decay  mainly into
fermion pairs since $M_h \lsim$~140~GeV.  This is, in general, also
the  dominant decay mode of the $A$ and $H$ bosons, since for $\tb \gg
1$,   they decay into $b \bar{b}$ and $\tau^+ \tau^-$ pairs with BRs
of the order of $\sim$ 90\% and 10\%, respectively. For large masses,
the top decay channels $H, A \rightarrow t\bar{t}$ open up, yet they
are suppressed  for large $\tb$. The $H$ boson can decay into gauge
bosons or $h$ boson pairs, and the $A$ particle into $hZ$ final
states; however, these decays are strongly suppressed for $\tb \gsim
5$.  The $H^\pm$ particles decay into fermions pairs: mainly
$t\bar{b}$ and $\tau \nu_{\tau}$ final states for $H^\pm$ masses,
respectively, above and below the $tb$ threshold.  If allowed
kinematically, they can also decay  into $hW$ final states for $\tb
\lsim 5$. Adding up the various decays, the widths of all five
Higgses remain  rather narrow; Fig.~\ref{Hdecay:MSSMw}.  

Other possible decay channels for the heavy $H, A$ and $H^\pm$
states, are decays into light charginos and neutralinos, which could
be important if  not dominant; decays of the $h$ boson into the
invisible lightest neutralinos (LSP) can also be important,
exceeding 50\% in some parts of the parameter space and altering the
searches at hadron colliders \cite{bbg4,SUSY-Filip}. Decays into
(third generation) squarks and sleptons can also occur for the heavy
Higgs states but are in general suppressed, in particular at high
$\tb$. See Ref.~\cite{Djouadi:2005gj} for more details. 

Note finally, that light SUSY particles can also affect the
branching ratio of the loop--induced modes in a sizable way
\cite{SUSYloops}. In particular, light stops can significantly
affect the $h \to gg$ decay mode while  light stops and charginos
can alter the photonic Higgs decays~\cite{SUSYloops,bbg4}.

\subsection{Higgs bosons in the CP-violating MSSM}\smallskip

A {\it quantitative} explanation of baryogenesis   requires physics
beyond the SM,  one possibility  being  an additional source of  CP
violation beyond the one present in the SM via CKM mixing. 
{CP--violation in the SUSY sector} is one such source which allows
explanation of baryogenesis at the electroweak scale. A general two
Higgs doublet model seems to be able to generate  adequate amount of
baryon asymmetry in the universe (BAU) and be consistent with the
current experimental constraints such as  electric dipole
moments~\cite{Huber:2006ri}.  In the MSSM, it may be possible to
satisfy all the low energy constraints and still have sufficient CP
violation in the theory to explain the BAU quantitatively (without
requiring too much fine-tuning, one needs to go to non-minimal
versions \cite{cline}). This further causes  new phases to enter the
MSSM Higgs sector, which is CP--conserving at tree--level, through
the large radiative  corrections. These phases affect both the
masses and the couplings of the neutral and charged Higgs particles,
thus having very serious implications for the Higgs phenomenology at
the LHC. This issue  has received a  lot of attention in the recent
times~\cite{cpnsh,mycpreview,CPHMSSM,HMSSMCPX,CPHmasses}. 

Since CP is violated, the three mass eigenstates $H_1,H_2,H_3$ 
need  no longer  have definite CP quantum  numbers and can be a
mixture of the $h,H,A$ states. The subscript $i$ indicates the order
of the mass $m_{H_i}$ of the $H_i$ boson in the spectrum, i.e.
$m_{H_1}\! <\! m_{H_2}\! <\! m_{H_3}$. It is obvious that this will
lead to significant  modification of the properties of the various
Higgs particles. Effect of this mixing on the couplings of  the
mixed CP states $H_1,H_2,H_3$ with a pair of gauge bosons/fermions
i.e., $H_i f \bar f$, $H_i V V$, can change the Higgs phenomenology
profoundly.  For details, see
e.g.~Refs.~\cite{cpnsh,mycpreview,CPHMSSM,HMSSMCPX,CPHmasses}. 

In multi-Higgs doublet models,  there exist sum rules
which force the different $H_i$ bosons to share among themselves the
coupling of the SM Higgs boson  to the massive gauge
bosons~\cite{mendez}, $\sum_i g_{H_iVV}^2 =g^2_{H_{\rm SM}}$. 
However, it is  only the CP--even component that is projected out. 
A CP violating MSSM is distinguished from a general CP violating
two-Higgs doublet model by the fact that the former has a prediction
for the mixing in terms of SUSY-breaking CP-violating phases of the
MSSM. The possible dilution of the LEP limits on the Higgs  masses
due to CP violation had been discussed in a model independent
formulation~\cite{gunion}.  The specific feature of the CP-violating
MSSM is the prediction for the mixing in terms of the SUSY
parameters and CP breaking phases that they have.

\begin{figure}[!h]
\vspace{-.7cm}
\begin{center}
\includegraphics[width=4.1cm,height=5cm]{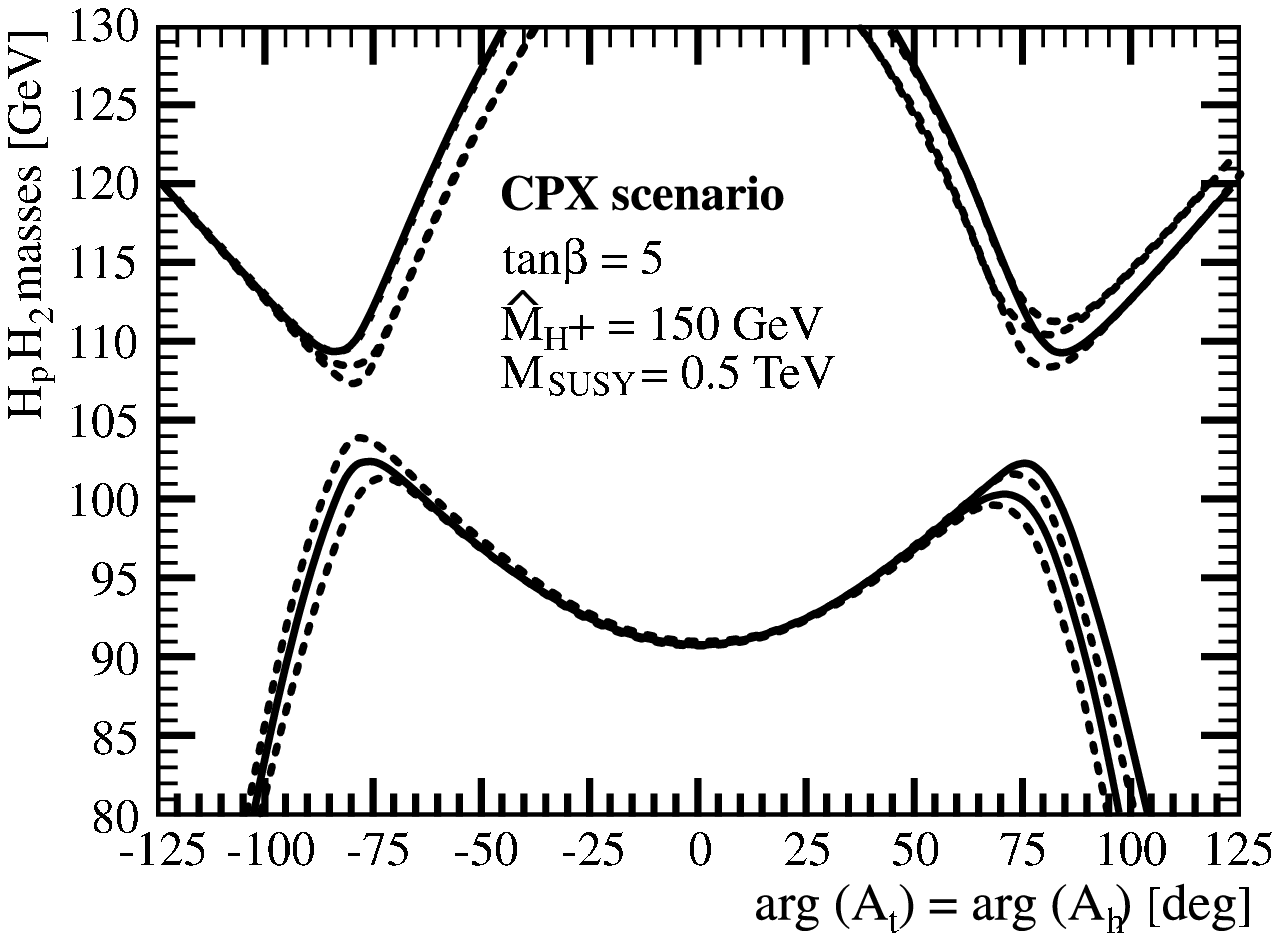}\hspace*{3mm}
\includegraphics*[width=4.2cm,height=5cm]{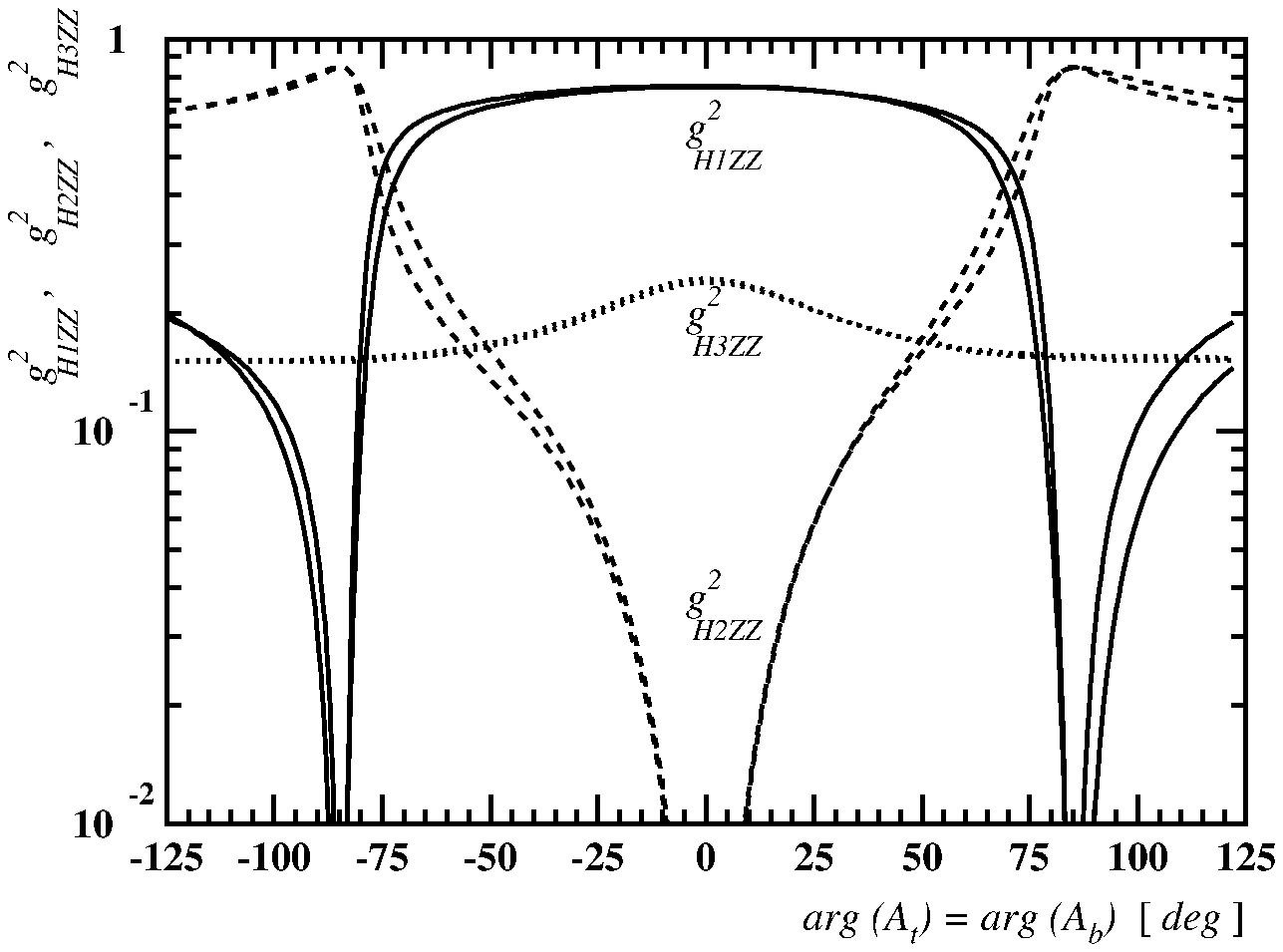}
\end{center}
\vspace*{-1cm}
\caption{The mass spectrum of neutral Higgs particles and their
couplings to the gauge bosons in the CP--violating MSSM CPX 
scenario (with Arg($A_t)$=Arg$(A_b)$=Arg($\mu)$ while Arg$(M_3)$=0
or $\frac{\pi}{2}$, all the other parameters are  
indicated on the figure) \protect\cite{CPHmasses}.}
\label{Hcoup:cpv}
\vspace*{-.7cm}
\end{figure}

As examples of new features in the CP violating MSSM, compared to
the usual MSSM, we simply mention the possibility of a relatively
light $H_1$ state with very weak couplings to the gauge bosons, and 
which could have escaped detection at LEP2 
\cite{CPHmasses,Abbiendi:2004ww,LEP2-HMSSM} and the possibility  of
resonant $H/A$ mixing when the two Higgs particles are degenerate in
mass \cite{HCPR1}. An example of the Higgs mass spectrum in the
so-called CPX scenario in which $H_1$ can be light  is shown in 
Fig.~\ref{Hcoup:cpv} (left) as a function of the phase of the
coupling $A_t$.

\begin{figure}[!h]
\vspace*{-7mm}
\centerline{
\includegraphics*[width=7cm,height=4.9cm]{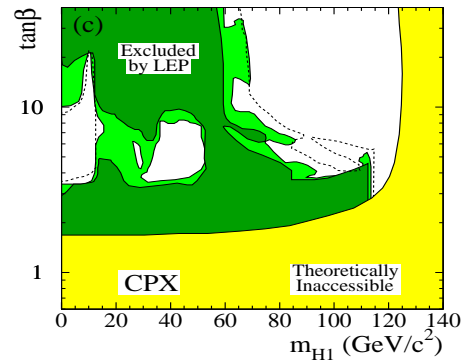}
}
\vspace*{-7mm}
\caption{Regions in the $\tan \beta$--$M_{H_1}$ plane disallowed
theoretically or excluded  by the current LEP searches 
\protect\cite{LEP2-HMSSM}. The allowed `hole' at the low $M_{H^+},
\tan\beta$ values can be seen very clearly.}
\label{opalexclusion}
\end{figure}

Fig.~\ref{Hcoup:cpv} (right) shows the result for the Higgs
couplings to gauge bosons in the same CPX scenario for two different
values of the gluino mass phase. In fact, the non observation of a
Higgs boson signal in the direct searches at the LEP, 
reinterpreted in the MSSM with CP violation, shows 
\cite{Abbiendi:2004ww,LEP2-HMSSM} that indeed there are
holes in the excluded region at small $\tan \beta$ and $m_{H_1}$ in
the $\tan \beta$--$M_{H_1}$ plane; see Fig.~\ref{opalexclusion}.  
This corresponds to the case of $H_1$ decoupled from the $Z$ boson,  
mentioned above.

Of course such features will have to be proved to be the result of 
CP--violation by, for instance,  studying CP--odd observables
and associated phenomenology.

\subsection{The Higgs sector in non-minimal SUSY}\smallskip

The Higgs sector in SUSY models can be more complicated than
previously discussed if some basic assumptions of the MSSM,  such
as  the presence of only two Higgs doublet fields, or R--parity
conservation, are relaxed.  A few examples are listed
below.\smallskip

\underline{The next--to--minimal SUSY extension, the NMSSM}, in which
the spectrum of the MSSM is extended by one singlet superfield
\cite{genNMSSM}, was among the first  SUSY models based on
supergravity-induced SUSY-breaking terms. It has gained a renewed
interest in the last decade, since it solves in a natural and elegant
way the so-called $\mu$ problem~\cite{MuProblem} of the MSSM; in the
NMSSM this parameter is linked to the vev of the singlet Higgs field,
generating a $\mu$ value close to the SUSY-breaking scale.
Furthermore, when the soft breaking terms are assumed to be
universal at the GUT scale, the resulting constrained model (cNMSSM)
is very constrained as one single parameter (e.g. the gaugino mass
$M_{1/2}$) allows to fully describe its phenomenology
\cite{cNMSSM}.  

\begin{figure}[h!]
\vspace*{-8mm}
\psfig{file=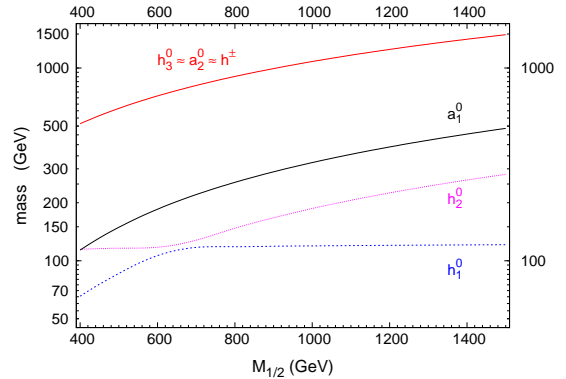,clip=,angle=270,
width=80mm} 
\vspace*{-9mm}
\caption{The Higgs masses as a function of the gaugino mass
parameter $M_{1/2}$ in the cNMSSM \cite{cNMSSM}.}
\label{fig:m12:phenoH}
\vspace*{-7mm}
\end{figure}

The NMSSM  leads to an interesting phenomenology
\cite{SUSY-NMSSM,benchmark} as the MSSM spectrum is extended to
include an additional  CP-even and CP-odd  Higgs states as well as a
fifth neutralino, the singlino. An example of the Higgs mass
spectrum in the cNMSSM \cite{cNMSSM} is shown in
Fig.~\ref{fig:m12:phenoH} as a function of the gaugino mass 
parameter. As in the MSSM in the decoupling regime, the heaviest
CP-even, CP-odd and charged Higgs states form a practically
degenerate SU(2)~multiplet with a common mass beyond 500 GeV; the
lightest CP-even state is mostly SM-like, with  a mass increasing
slightly with $M_{1/2}$ from 115~GeV up to $\sim 120$~GeV.   The
third CP-even state has a dominant singlet component:  for small
$M_{1/2}$ it is lighter than the SM-like Higgs boson, escaping LEP
constraints due to the very small coupling to the $Z$~boson. For
increasing values of $M_{1/2}$, its mass increases until it becomes
comparable and eventually  exceeds  the mass of  SM-like CP-even
Higgs state.

However, in the unconstrained NMSSM, the effect of the additional 
singlet to  the scalar potential leads to a relaxation of the  upper
bound on the mass of the lighter CP--even  particle above that of
the MSSM $h$ boson~\cite{NMSSM-bound}. Further, the constraints 
in the $M_A$--$\tb$ plane, implied by the negative results of the
LEP2  searches are less restrictive as compared to those in the 
MSSM~\cite{cpnsh,dpdrees}. In addition,  there exists a small
region  not yet  completely excluded, where the lightest CP--even
Higgs boson might have escaped the LEP2 searches.  Even more
interestingly, there exist possibilities that one of the neutral
Higgs particles, in particular the lightest  pseudoscalar $A_1$, is
very light with a mass of a few ten's of GeV.  The  light CP--even
Higgs, which is SM--like in general, could then  decay into pairs of
$A_1$ bosons, $H_1 \to A_1 A_1 \to 4b, 4\tau$, with  a large
branching fraction.\smallskip

\underline{Higgs bosons in GUT theories.} A large variety of
theories,  string theories, grand unified theories,  left--right
symmetric models, etc., suggest an additional gauge symmetry which
may be broken only at the TeV scale.   This leads to an extended
particle spectrum and, in particular, to additional Higgs fields
beyond the minimal set of the MSSM. Especially common are new U(1)'
symmetries broken by the vev of a singlet field (as in the NMSSM) 
which leads to the presence of a $Z'$ boson and one additional
CP--even Higgs particle compared to the MSSM; this is the  case, for
instance, in the exceptional MSSM \cite{H-GUT-ESSM} based on the
string inspired $E_6$ symmetry. The secluded ${\rm SU(2)\times U(1)
\times U(1)'}$ model \cite{H-GUT-secluded}, in turn, includes four 
additional singlets that are charged under U(1)', leading to  6
CP--even and 4 CP--odd neutral Higgs states. Other exotic Higgs
sectors \cite{cpnsh,H:higheR} are, for instance, Higgs
representations that transform as SU(2) triplets or  bi--doublets
under the ${\rm SU(2)_L}$ and  ${\rm SU(2)_R}$ groups in left--right
symmetric models, that are motivated by the seesaw approach to 
explain the small neutrino masses and which lead e.g. to a doubly
charged Higgs  boson $H^{--}$. These extensions, which also predict
extra matter fields, would lead to a very interesting phenomenology
and new collider signatures in the Higgs sector. We will not be
discussing much about this subject in this review.

\begin{figure}[!h]
\begin{center}
\mbox{
\includegraphics[width=7cm,height=4.5cm]{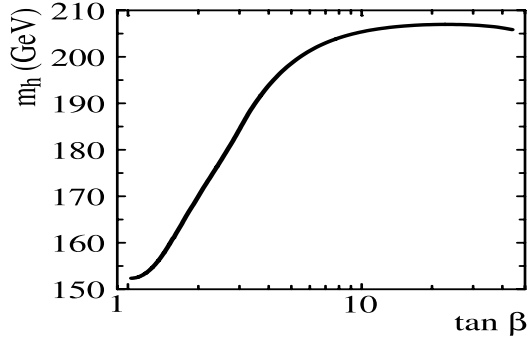}
}
\end{center}
\vspace*{-1.2cm}
\caption
{The upper bound on the lighter Higgs mass in a general SUSY model
\cite{HSUSY-thbound}.}
\label{Hmass:nmssm}
\vspace*{-7mm}
\end{figure}

In a general SUSY model, with an arbitrary number of singlet and
doublet scalar fields (as well as a matter content which  allows for
the unification of the gauge couplings), a linear combination of
Higgs fields has to generate  the $W/Z$ masses and thus, from the
triviality argument discussed earlier, a Higgs particle should have
a mass below 200 GeV  and significant couplings to gauge bosons
\cite{HSUSY-thbound}. The upper bound on the mass of the lightest
Higgs boson in this most general SUSY model is displayed in 
Fig.~\ref{Hmass:nmssm} as a function of $\tb$. \smallskip

\underline{R--parity violating models.}    Models in which 
R--parity is spontaneously broken [and where one needs to either
enlarge the SM symmetry or the spectrum to include additional gauge
singlets],  allow for an explanation of the light neutrino data
\cite{H-RparityV}.  Since $\not \hspace*{-1.5mm}R_p$ entails the
breaking of the total lepton number $L$, one of the CP--odd scalars,
the Majoron $J$, remains massless being  the Goldstone boson
associated to $\not \hspace*{-1.5mm}L$. In these models, the neutral
Higgs particles have also reduced couplings to the gauge bosons.
More importantly,  the CP--even Higgs particles can decay into pairs
of invisible Majorons, $H_i \to JJ$, while the CP--odd particle can
decay into a CP--even Higgs and a Majoron, $A_i \to H_i J$, and
three Majorons,  $A \to JJJ$ \cite{H-RparityV}.

\subsection{Higgs bosons in alternative scenarios}\smallskip

There are also many non supersymmetric extensions of the SM which might lead to
a different Higgs phenomenology. In some cases, the Higgs sector would consist
of one scalar doublet leading to a Higgs boson which would mimic the SM Higgs,
but the new particles that are present in the models might alter some of its
properties. In other cases, the Higgs sector is extended to contain  additional
scalar fields leading to the presence of new Higgs particles. Another
possibility is a scenario with a composite and strongly interacting Higgs,  or
where no Higgs particle is present at all, leading to strong interactions of  
the $W/Z$ bosons.  Below we  give a non exhaustive list of various possible 
scenarios.\smallskip

\begin{figure}[h!]
\begin{minipage}{7cm}
\includegraphics[width=7cm,height=6cm]{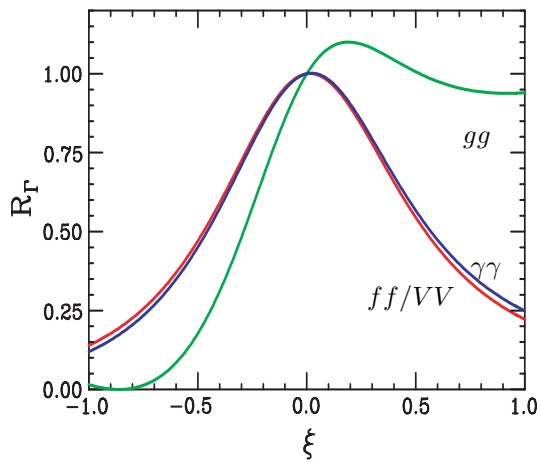}
\end{minipage}
\hspace*{-2cm}
\begin{minipage}{1.cm}\hspace*{.7cm} $gg$ \vspace*{.9cm}

\hspace*{.8cm} $\gamma \gamma$\\
\hspace*{-.5cm}$ff/VV$\\
\end{minipage}
\vspace*{-6mm}
\caption{The ratio $R_\Gamma$ of Higgs partial widths to their SM 
values, as a function of the Higgs-radion mixing parameter $\xi$ 
with $M_H=125$ GeV, $M_\phi=300$ GeV and a scale $v/\Lambda=0.2$ 
\cite{Hewett:2002nk}.}
\label{fig:Hradion}
\vspace*{-7mm}
\end{figure}

\underline{Scenarios with Higgs mixing.}  In warped extra--dimensional
models \cite{WED} the fluctuations of the size of the extra  dimension
about its stabilized value manifest themselves as a single scalar 
field, the radion. In the Randall Sundrum model with a bulk scalar
field,  it is expected that  the radion is the lightest state beyond
the SM fields with a mass probably  in  the range between ${\cal
O}$(10 GeV)  and $\Lambda={\cal O}$(TeV)
\cite{Hewett:2002nk,Chaichian:2001rq,Dominici:2002jv}. The couplings of 
the radion are
order of  $1/\Lambda$ and are very similar to the couplings of the SM
Higgs boson,  except  for one important difference: due to the trace
anomaly, the radion directly  couples to massless gauge bosons at one
loop.  Moreover, in the low energy four--dimensional effective theory,
the radion can mix with the  Higgs boson.  This mixing  can lead to
important  shifts in the Higgs couplings which become apparent in the
Higgs decay widths and production cross sections;
Fig.~\ref{fig:Hradion}.

\begin{figure}[h!]
\begin{minipage}{5cm}
\hspace*{8mm}
\includegraphics[width=5.2cm,height=6cm]{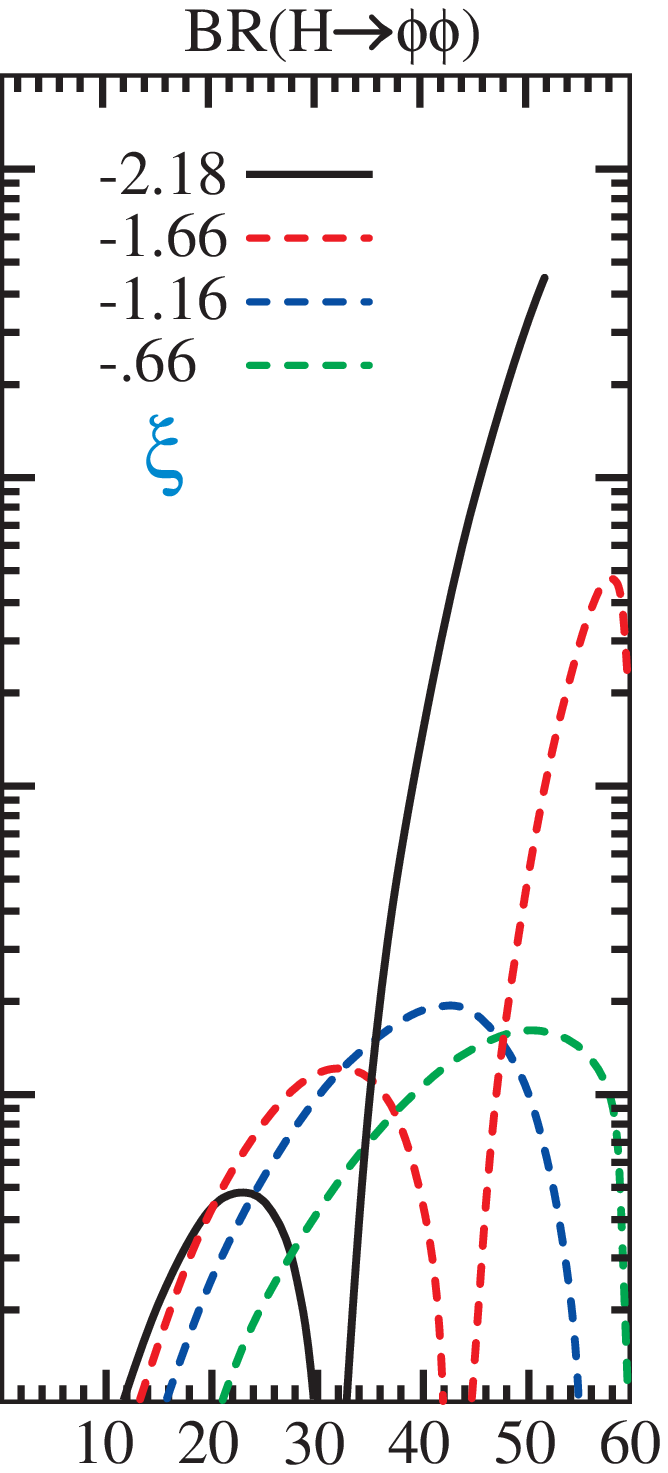}
\end{minipage}
\hspace*{.7cm}
\begin{minipage}{1.5cm}
\hspace*{1.5cm} \\  \vspace*{.7cm} $1$\\ \vspace*{.7cm} $10^{-1}$\\  \vspace*{.7cm} 
 $10^{-2}$\\  \vspace*{.7cm} $10^{-3}$\\  \vspace*{.7cm} $10^{-4}$
 \hspace*{-1cm}
\end{minipage}
\vspace*{-8mm}
\caption{The branching fractions for the decays $H\to \phi \phi$ as 
a function of $M_\phi$ for different $\xi$ values and $M_H=120$ GeV, 
$\Lambda=5$ TeV \cite{Dominici:2002jv}.}
\label{fig:HradionI}
\vspace*{-8mm}
\end{figure}

Another important consequence of radion mixing is the decays of the
Higgs boson into a pair of radions. Indeed, if the radion is
relatively light, the decays $H\to \phi \phi$ might be kinematically
accessible and, for some mixing values, the branching fractions
might be substantial. This is exemplified in Fig.~\ref{fig:HradionI}
where BR($H\to \phi\phi$)  is displayed  in a specific scenario.

In large extra dimension models \cite{LED}, mixing of the Higgs boson
with graviscalars  also occurs \cite{H-graviscalars}, leading to an
invisible decay width.  Mixing effects also occur if the SM is
minimally extended in a renormalizable way to contain a singlet scalar
field $S$ that does not couple to the other SM particles; its main
effect would be to alter the scalar potential and to  mix with the SM
Higgs field \cite{NMSM} and, in such a case, the Higgs could mainly
decay into two invisible $S$ particles.\smallskip

\underline{Scenarios with extended Higgs/gauge/matter.}
Non--supersymmetric  extensions of the Higgs sector with additional
singlet, doublet and higher representation fields have also been
advocated \cite{H:higheR}. Examples are the minimal SM extension
with a singlet discussed above, two--Higgs doublet models which
potentially include CP--violation, triplet Higgs fields in models
for light neutrino mass generation, etc...  These extensions lead
to a rich spectrum of Higgs particles which could be produced at
the LHC. In other extensions of the SM, new gauge bosons and new
matter particles are predicted and they can affect the properties
of the SM--like Higgs boson. For instance the new fermions present
in little Higgs and extra--dimensional  models might contribute to
the loop induced Higgs couplings, while new heavy gauge bosons
could alter the Higgs couplings to $W$ and $Z$ bosons for
instance. The anomalous $ZZH$ and $t \bar t H$ couplings can be a good probe 
of the additional scalars and/or the novel features of the geometry in
the extra dimensions~\cite{ZZHED}. \smallskip

\underline{Scenarios with a composite Higgs boson.} In little Higgs
models \cite{LHM},  the dynamical scale is around $\Lambda=10$ TeV,
unlike the traditional Technicolor model
\cite{H-technicolor}.  A light Higgs boson can be
generated as a pseudo Goldstone boson  and its mass  of order 100
GeV is protected against  large radiative corrections individually
in the boson and the fermion sectors. The models predict a rich
spectrum of new particles not only at the scale $\Lambda$ but also
at lower scales. Axion--type pseudoscalar bosons may be associated
with the spontaneous breaking of U(1) factors in the extra global
symmetries \cite{Kilian:2006eh}. These particles have properties
analogous to Higgs bosons and  can be produced at the LHC;
deviations in the production and decay rates of the  SM--like Higgs
boson can also be induced by these particles. Note that, recently,
a model--independent description of a strongly  interacting light
Higgs  has been given \cite{H-SILH}.\smallskip  

\underline{Higgsless models and strong $W/Z$ interactions.}  
Assuming the $W/Z$ bosons to become  strongly interacting at TeV
energies,  damping the rise of the elastic $W/Z$ scattering 
amplitudes, is an alternative way to solve the problem of unitarity
violation at high energies in the SM, without adding a relatively 
light Higgs boson.
Naturally, the strong forces between the massive gauge
bosons may be traced back to new fundamental interactions
characterized by a scale of order 1 TeV \cite{H-technicolor}. Also
in theories with extra space dimensions, EWSB can occur  without
introducing additional fundamental scalar fields, leading  also to
Higgsless theories \cite{Hless}. Studying such difficult scenarios
at the LHC will be possible with very high luminosity
\cite{HlessLHC}.

\section{Higgs production and detection at the LHC}

\subsection{The SM Higgs case}\smallskip

\begin{figure}[!h]
\vspace*{-.8cm}
\begin{center}
\vspace*{-1.cm}
\hspace*{-1.5cm}
\includegraphics*[width=9cm,height=6cm] {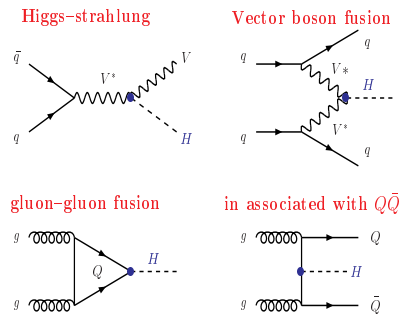}
\end{center}
\vspace*{-1.3cm}
\caption{The production mechanisms for SM Higgs bosons at hadron colliders.}
\label{ppmecha-lhc}
\vspace*{-.5cm}
\end{figure}

There are essentially four mechanisms for the single production of
the SM Higgs boson at hadron colliders \cite{P1}; some Feynman
diagrams are shown in  Fig.~\ref{ppmecha-lhc}.  

\begin{figure}[!h]
\vspace*{-.8cm}
\begin{center}
\epsfig{file=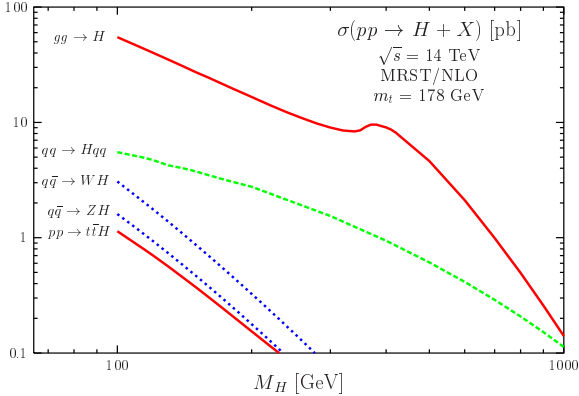,width=7.5cm,height=6.0cm} 
\end{center}
\vspace*{-1.2cm}
\caption{The production cross sections for the SM Higgs boson at the 
LHC in the main channels.}
\label{ppproduct-lhc}
\vspace*{-.5cm}
\end{figure}

The total cross sections, obtained with the  programs of
Ref.~\cite{Michael}, are displayed in Fig.~\ref{ppproduct-lhc} for 
the LHC with $\sqrt{s}=14$ TeV as a function of the Higgs mass; the
top quark mass is set to $m_t=178$ GeV and the MRST parton
distributions functions \cite{MRST} have been adopted. The NLO, and
eventually NNLO, corrections have been implemented as will be
summarized below, where we discuss the main features of each
production channel.\smallskip

{\bf a)} \underline{$gg \to H$}: This is by far the dominant
production process at the LHC, up to masses $M_H \approx 1$ TeV. The
most promising detection channels are \cite{gg-detection} $H \to
\gamma \gamma$ for $M_H \lsim 130$ GeV and slightly above this mass
value, $H\to ZZ^* \to 4\ell^\pm$ and $H\to WW^{(*)}\to \ell \ell \nu
\nu$ with $\ell=e,\mu$ for masses below, respectively, $2M_W$ and
$2M_Z$. For higher Higgs masses, $M_H \gsim 2M_Z$, it is the golden
mode $H \ra ZZ \ra 4\ell^\pm$, which from $M_H \gsim 500$ GeV can be
complemented by $H \to ZZ \to \nu\bar{\nu} \ell^+ \ell^-$ and $H \to
WW \to \nu \ell jj$ to increase the statistics 
\cite{atlastdr,CMSTDR,LHC,Houches,Houches-last,Weiglein:2004hn}.  

\begin{figure}[!h]
\begin{center}
\includegraphics*[width=8.5cm,height=7cm]{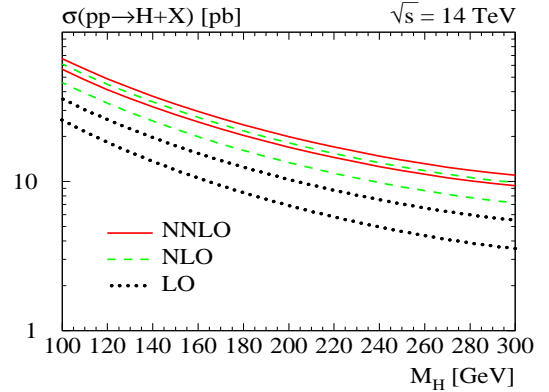} 
\end{center}
\vspace*{-1.9cm}
\caption{SM Higgs production cross sections in the $gg$ fusion 
process at the LHC as a function  of $M_H$ at the three different orders
with the upper (lower) curves  are for the choice of the 
renormalization and factorization scales $\mu=\frac{1} {2} M_H$ ($2M_H$); 
from Harlander and Kilgore in Ref.~\cite{P3}.} 
\label{Hcsec:nnlo}
\vspace*{-.5cm}
\end{figure}

The next--to--leading order (NLO) QCD corrections have been
calculated in both the limit where the internal top quark has been
integrated out \cite{P2inf}, an approximation which should be valid
in the Higgs mass range $M_H \lsim 300$ GeV, and in the case where
the full quark mass dependence has  been taken into account
\cite{P2}. The corrections lead to an increase of the  cross
sections by a factor of $\sim 1.7$. The ``tour de force" of deriving
the three--loop corrections has been preformed in the infinite
top--quark mass limit; these NNLO corrections lead to the increase
of the rate by an additional 30\% \cite{P3} 
(see also Refs.~\cite{P3-add,P3-add2}. This results in a nice convergence 
of the perturbative series and a strong reduction  of the scale
uncertainty, which is the measure of unknown higher order effects;
see  Fig.~\ref{Hcsec:nnlo}. The resummation of  the soft and
collinear corrections, performed at next--to--next--to--leading
logarithm accuracy,  leads to another increase of the rate by $\sim
5\%$ and a decrease of the scale uncertainty \cite{SG-resum}.  The
QCD corrections to the differential  distributions, and in
particular to the Higgs transverse momentum and rapidity
distributions, have also been recently calculated at NLO [with a
resummation for the former] and shown to be rather large
\cite{Pt-eta-distrib}. The dominant components of the electroweak
corrections, some of which have been derived very recently, are
comparatively very  small \cite{EW-CR}.\smallskip  

{\bf b)} \underline{$q\bar q \to HV$}: The associated production
with gauge bosons, with $H \to b\bar{b}$ and possibly $H \to WW^*
\to \ell^+ \nu jj$, is the most relevant mechanism at the Tevatron
\cite{tevreview} [$gg \to H \to W W \to \ell \nu \ell \nu$ being
important for Higgs masses close to 160 GeV]. At the LHC, this process
plays only a marginal role; however, the channels $HW \to \ell \nu
\gamma \gamma$ and eventually $ \ell \nu b\bar b$ could be useful
for the measurement of Higgs couplings.

\begin{figure}[!h]
\vspace*{-.7cm}
\begin{center}
\includegraphics*[width=7.9cm]{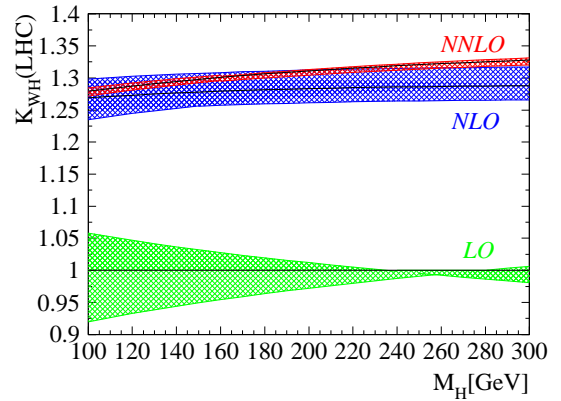}
\end{center}
\vspace*{-1.4cm}
\caption{ $K$-factors for $pp \to  HW$ at the LHC as a function of
$M_H$ at LO, NLO and NNLO with the bands  represent the spread of the
cross section when the scales are varied in  the  range $\frac{1}{3}
M_{HV} \leq \mu_R\, (\mu_F) \leq 3M_{HV}$ \cite{HV-NNLO}.}
\label{Hcsec:kfactor}
\vspace*{-.6cm}
\end{figure}

The QCD corrections, which at NLO \cite{HV-NLO,HV+VV-NLO}, can be
inferred from Drell--Yan production, have been calculated at NNLO
\cite{HV-NNLO}; they are $\sim$  30\% in total; see
Fig.~\ref{Hcsec:kfactor}. The ${\cal O} (\alpha)$ electroweak
corrections have been also derived recently \cite{HV-EW} and
decrease the rate by $5$ to 10\%. The remaining scale dependence is
very small, making this process the theoretically cleanest of all
Higgs production processes.\smallskip

\begin{figure}[!h] 
\vspace{-0.5cm}
\includegraphics*[scale=0.37,angle=90]{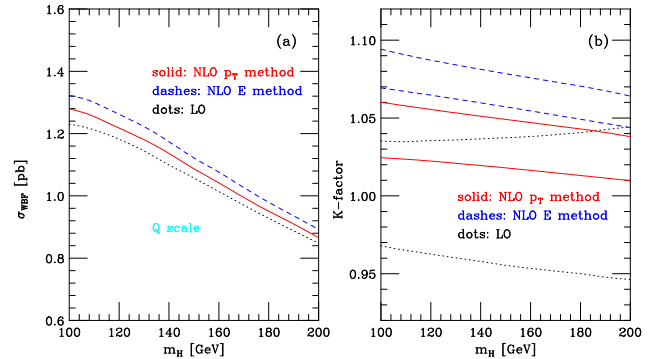}
\vspace*{-1.cm}
\caption{The $pp\to Hqq$ cross section after cuts as a function of 
$M_H$ at LO (dotted line) and NLO with the tagging jets defined in the $P_T$ 
and $E_T$ methods (left) and the scale variation of the LO and NLO cross 
sections as a function of $M_H$ (right); from Ref.~\cite{VV-MC}.}
\label{Hsec:hqqnlo}
\vspace*{-.5cm}
\end{figure}

{\bf c)} \underline{The $WW/ZZ$ fusion mechanism}: This process  has
the second largest cross section at the LHC. The QCD corrections,
which can be obtained in the structure--function approach, are at
the level of 10\% and thus small \cite{HV+VV-NLO,VV-NLO} (the
electroweak corrections \cite{VV-NLO-EW} are at the level of a few
percent). The corrections including cuts, and in particular
corrections to the $p_T$ and $\eta$ distributions, have also been
calculated  and implemented into a parton--level Monte--Carlo
program \cite{VV-MC}. With the specific cuts to the process, the
output for the production cross section is shown in
Fig.~\ref{Hsec:hqqnlo} for a Higgs  in the mass range 100--200 GeV. 

For several reasons, the interest in this process has grown in
recent years: it has a large enough cross section [a few picobarns
for $M_H \lsim 250$ GeV] and one can use cuts, forward--jet tagging,
mini--jet veto for low luminosity as well as triggering on the
central Higgs decay products] \cite{WWfusion0}, which render the
backgrounds comparable to the signal, therefore allowing precision
measurements. In the past, it has been shown that the decay $H \to
\tau^+ \tau^-$ and possibly $H \to \gamma \gamma , ZZ^*$ can be
detected and could allow for coupling measurements
\cite{Houches,Dieter,Dieter1}.  In the last years, parton--level
analyzes have shown that various  other channels can be possibly
detected \cite{WWfusion,WWinvis}:  $H \to WW^*$ for $M_H \sim$ 125--180 GeV,
$H \to \mu^+ \mu^-$ [for  second--generation coupling measurements],
$H \to b\bar{b}$ [for the $b\bar{b}H$ Yukawa coupling] and $H \to $
invisible (see later). Recent experimental  simulations \cite{LHC}
have assessed more firmly the potential of this channel. \smallskip 

\begin{figure}[!h]
\vspace{-.7cm}
\includegraphics[width=7cm,height=5cm]{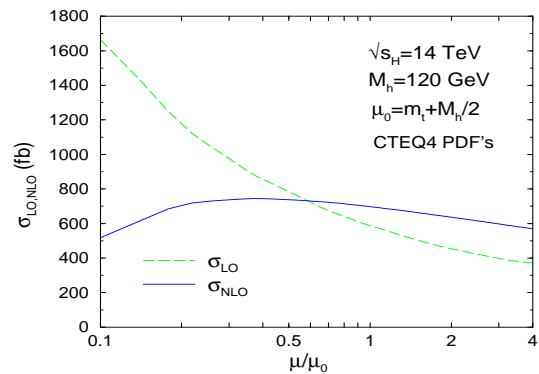}
\vspace*{-.7cm}
\caption{The Higgs production cross sections in  the $t\bar{t}H$ process as a 
function  of the renormalization/factorization scale $\mu$; from 
Dawson et al.~\cite{ttH}.}  
\label{Hcsec:ttbarH}
\vspace*{-.5cm}
\end{figure}

{\bf d)} \underline{$pp \to t\bar t H$:} Finally, Higgs boson
production in association with top quarks, with $H \to \gamma
\gamma$ or $b\bar{b}$, can in principle be observed at the  LHC and
direct measurement of the top Yukawa coupling, as well as an 
unambiguous determination of the CP of the Higgs can be possible.  
(Recent analyses have however, shown that $pp \to t\bar t H\to t\bar
t b\bar b$ might be subject to a too large jet background
\cite{CMSTDR}).  The cross  section is rather involved at
tree--level since it is a three--body process, and the calculation
of the NLO corrections was a real challenge which was met  a few
years ago \cite{ttH}. The $K$--factors turned out to be rather
small, $K\sim 1.2$ at the LHC. However, the scale dependence is
drastically reduced from a factor two at LO to the level of 10--20\%
at NLO; see  Fig.\ref{Hcsec:ttbarH}. Note that the NLO corrections
to the $q\bar q /gg \to b\bar b H$  process, which is more relevant
in the MSSM, have been also completed \cite{bbH}: compared with the
NLO rate for the $bg \to bH$ process where the  initial $b$-quark is
treated as a parton \cite{bg-bH,Kfac-H+}, the calculations agree  within the
scale uncertainties \cite{bbH-comp}. A similar situation
occur for $H^\pm$ production in the $gb$ process: the $K$--factor is
moderate $\sim\!1.2$--1.5 if the cross section is evaluated at
scales $\mu \sim \frac12 (m_t+M_{H^\pm})$ \cite{Kfac-H+}. 

Note that the PDF uncertainties have also been  estimated for the four
production processes:  at the LHC, the uncertainties range from 5\% to
15\% depending on the considered process and the Higgs mass
\cite{Samir-PDFs}.  

\begin{figure}[!h]
\includegraphics*[width=8cm,height=5cm]{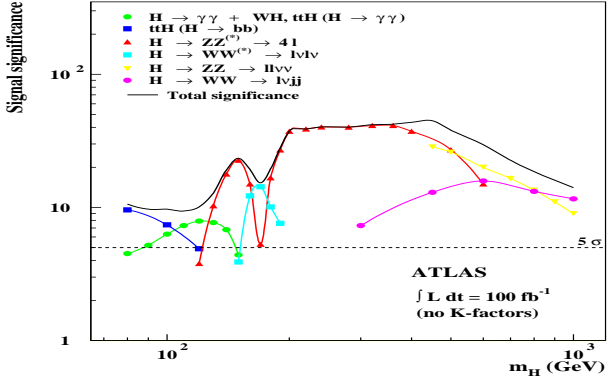}
\vspace*{-.8cm}
\caption{Significance for the experimental detection
\protect\cite{atlastdr} of the SM Higgs boson at the LHC.}
\label{Hdetect:ATLAS}
\vspace*{-.6cm}
\end{figure}

All the various channels discussed above have been discussed in great detail
over the past 
decades~\cite{atlastdr,CMSTDR,LHC,Houches,Houches-last,Weiglein:2004hn}. The
significance for detecting the SM Higgs particle in the various production and
decay channels is shown in  Fig.~\ref{Hdetect:ATLAS}, assuming a 100
fb$^{-1}$ integrated luminosity. 
 
\subsection{The CP conserving MSSM }
\vspace*{1mm}


In the CP conserving MSSM, the production processes for the CP--even
$h,H$ bosons are practically the same as for the SM Higgs and the
ones depicted in Fig.~\ref{ppmecha-lhc} are all relevant. However,
the $b$ quark will play an important role for moderate to large
$\tb$ values as its Higgs couplings are enhanced.  First, one has to
take into account the $b$ loop contribution  in the $gg \to h,H$ 
process which becomes the dominant component in the MSSM [here, the 
QCD corrections are available only at NLO  where they have been
calculated  in the full massive case \cite{P2} and increase the rate
by $\sim 1.5$; SUSY-QCD corrections are discussed in 
Refs.~\cite{wess-rev,SQCD-rev} e.g.].  Moreover, in associated Higgs
production with heavy quarks, $b\bar{b}$ final states must be
considered, $pp \to b \bar b + h/H$, and this process for either $h$
or $H$ becomes the dominant one in the MSSM [here, the QCD
corrections are available in both the $gg$ and $gb \to b\Phi, b\bar
b \to \Phi$ pictures \cite{bbH,bg-bH,Kfac-H+,bbH-comp} depending on how many
$b$--quarks are to be tagged, and which are equivalent if the
renormalization and factorization scales are chosen to be small, 
$\mu  \sim \frac14 M_\Phi$]. The rates for  associated production
with $t\bar{t}$ and $W/Z$ pairs as well as for $WW/ZZ$ fusion
processes, are suppressed for at least one of the particles as a
result of coupling reduction.

Because of CP invariance which forbids $AVV$ couplings, the $A$
boson cannot be produced in the Higgs-strahlung and vector boson
fusion processes; the rate  for the $pp \to t\bar t A$ process is
suppressed by the small $At\bar t$ couplings for $\tb \gsim 3$.
Hence, only the $gg\to A$ fusion with the $b$--quark loops included
[where the QCD corrections are also available only at NLO and
are approximately the same as for the CP--even Higgs boson with
enhanced  $b$--quark couplings] and associated production with
$b\bar b$ pairs, $pp \to b \bar b +A$ [where the QCD corrections are
the same as for one of the CP--even Higgs bosons as a result of
chiral symmetry] provide large cross sections.  The
one--loop induced processes  $gg \to AZ, gg\to Ag$ [which hold also
for one of the CP--even Higgses] and associated production with
other Higgs particles, $pp \to A+h/H/H^+$  are possible but the
rates are much smaller in general, in particular for $M_A \gsim 200$
GeV \cite{gg-others}.

For the charged Higgs boson, the dominant channel is the production
from top quark decays, $t \to H^+ b$, for masses not too close to
$M_{H^\pm}=m_t\!-\! m_b$; this is particularly true at low or large
$\tb$ when the $t\to H^+b$ branching ratio is significant. For
higher masses \cite{pp-H+},  the processes to be considered is the
fusion process $gg \to H^\pm tb$ supplemented by $gb \to H^\pm t$.
The two processes have to be properly combined and the NLO
corrections  for both  processes have been derived \cite{Kfac-H+}
and are moderate, increasing the cross sections by 20 to 50\%  if
they are  evaluated at low scales, $\mu \sim \frac12
(m_t+M_{H^\pm})$.  Additional sources \cite{pp-H+-others} of
$H^\pm$ states for masses below $M_{H^\pm} \approx 250$ GeV are
provided by pair and associated production with neutral Higgs bosons
in $q\bar q$ annihilation as well as $H^+H^-$ pair and associated
$H^\pm W^\mp$ production in $gg$ and/or $b\bar b$ fusion but the
cross sections are not as large,  in particular for $M_{H^\pm} \gsim
m_t$.

\begin{figure}[!h]
\vspace*{-8.mm}
\begin{center}
\includegraphics*[width=8.7cm,height=6.1cm]{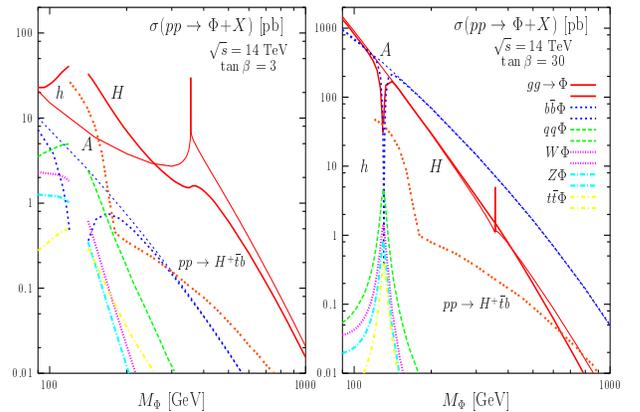}
\end{center}
\vspace{-1.3cm}
\caption{The cross section for the neutral and charged MSSM Higgs production in
the main channels at the LHC as a function of their respective masses for
$\tb=3$ and $30$ in the maximal mixing scenario. 
}
\label{Hcsec:mssmlhc}
\vspace*{-.1cm}
\end{figure}

The cross sections for the dominant production mechanisms are shown
in  Fig.~\ref{Hcsec:mssmlhc}.  as a function of the Higgs masses for
$\tb=3$ and $30$ for the same set of  input parameters as used for the
widths and branching ratios. The NLO QCD  corrections are included,
except  for the $pp \to Q \bar Q\,$Higgs processes where, however, the
scales have  been chosen as to approach the NLO results;  the MRST NLO
structure functions have been adopted.  As can be seen, at high $\tb$,
the largest cross sections are by far those of the $gg \to \Phi_A/A$
and $q\bar q/ gg \to b\bar b+ \Phi_A/A$ processes, where $\Phi_A=H\,
(h)$ in the (anti--)decoupling regimes $M_A > (<) M_h^{\rm max}$: the
other processes involving these two Higgs bosons have cross sections
that are several orders of magnitude smaller. The production cross
sections for the other CP--even Higgs boson, that is $\Phi_H=h\,(H)$
in the (anti--)decoupling regime when $M_{\Phi_H} \simeq M_h^{\rm
max}$, are similar to those of the SM Higgs boson with the same mass
and are substantial in all the channels which have been displayed. At
small $\tb$, the $gg$ fusion and $b\bar b$--Higgs cross sections are
not strongly enhanced as before and all production channels, except
for $b\bar b$--Higgs which is only slightly enhanced, have cross
sections that are smaller than in the SM Higgs case, except for $h$ 
in the decoupling regime.

\begin{figure}[!h]
\begin{center}
\includegraphics[width=7.cm,height=6cm]{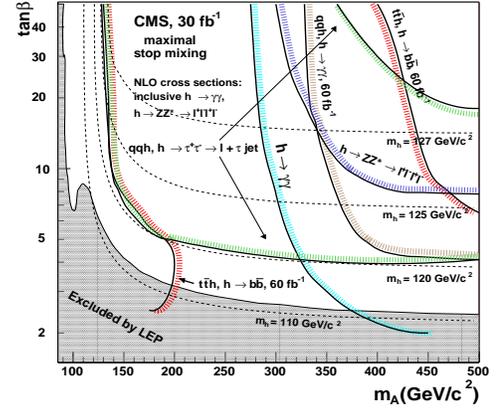}
\end{center}
\vspace*{-.9cm}
\caption{The areas in the $(M_A, \tb)$ parameter space where the 
lighter MSSM Higgs boson can be discovered at the LHC with
an integrated luminosity of 30 fb$^{-1}$ in the standard production 
channels \cite{CMSTDR}.
}
\label{Hdetec:mssmlhch}
\vspace*{-.7cm}
\end{figure}

\begin{figure}[!h]
\vspace*{-.5cm}
\begin{center}
\includegraphics[width=7.cm,height=5.8cm]{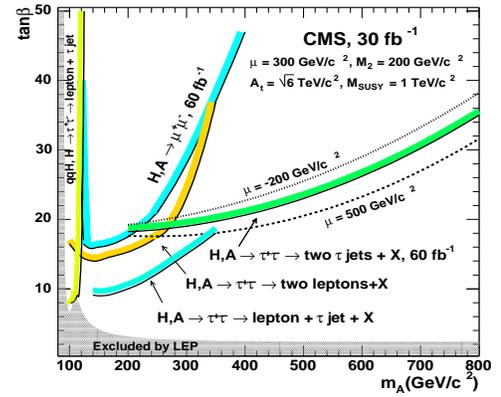}
\end{center}
\vspace*{-1.1cm}
\caption{The areas in the $(M_A, \tb)$ parameter space where the 
heavier MSSM neutral Higgs bosons can be discovered at the LHC with
a luminosity of 30 fb$^{-1}$ in the standard production 
channels \cite{CMSTDR}.
}
\label{Hdetec:mssmlhcHA}
\vspace*{-.7cm}
\end{figure}

The principal detection signals of the neutral Higgs bosons at the LHC, in the
various regimes of the MSSM, are as follows
\cite{Djouadi:2005gj,atlastdr,CMSTDR,LHC,Houches,intense}.  

a) \underline{Decoupling regime}: One of the most interesting region
is the decoupling regime, i.e. when $M_h \simeq M_h^{\rm max}$, the
lighter $h$ boson is SM--like and has a mass smaller than $\approx
140 $ GeV. It can be detected in the $h \to \gamma \gamma$ decays
[possibly supplemented with a lepton in associated $Wh$ and $t\bar t
h$ production], and eventually in $h\to ZZ^*, WW^*$ decays in the
upper mass range, and if the vector boson fusion processes are used,
also in the decays $h \to \tau^+ \tau^-$ and eventually $h \to W
W^*$ in the higher mass range $M_{h} \gsim 130$ GeV; see
Fig.~\ref{Hdetec:mssmlhch}.  For relatively large values of $\tb$
$(\tb \gsim 10)$, the heavier CP--even $H$ boson which has enhanced
couplings to down--type fermions, as well as the pseudoscalar Higgs
particle, can be observed in the process $pp \to b\bar b + H/A$
where at least one $b$--jet is tagged and with the Higgs boson
decaying into $\tau^+ \tau^-$, and eventually, $\mu^+ \mu^-$ pairs
in the low mass range. With a luminosity of 30 fb$^{-1}$ (and in
some cases lower) a large part of the $[\tb,M_A]$ space can be
covered  as can be seen from Fig.~\ref{Hdetec:mssmlhcHA}.  

b) \underline{Anti-decoupling regime}: In the anti-decoupling regime,
i.e. when $M_A < M_h^{\rm max}$ and at high $\tb$ ($\gsim 10$), it is
the heavier $H$ boson which will be SM--like and can be detected as
above, while the $h$ boson will behave like the pseudoscalar Higgs
particle and can be observed in $pp \to b\bar b+ h$ with $h \to \tau^+
\tau^-$ or $\mu^+ \mu^-$ provided its mass is not too close to $M_Z$
not to be swamped by the background from $Z$ production.  The part of
the $[\tb,M_A]$ space which can be covered is also shown in 
Fig.~\ref{Hdetec:mssmlhcHA} and corresponds to $M_A \lsim 130$ GeV.

\begin{figure}[!h]
\vspace*{-.5cm}
\begin{center}
\includegraphics[width=6 cm]{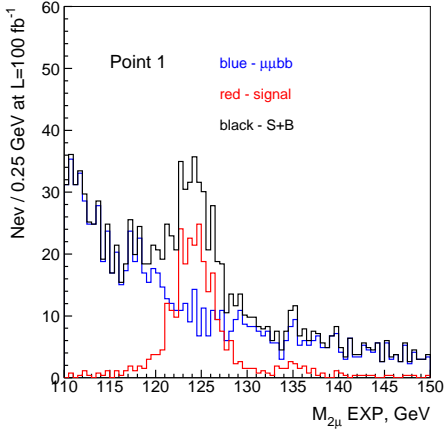}
\end{center}
\vspace*{-1.1cm}
\caption{The $\mu^+ \mu^-$ pair invariant mass distributions for the 
three Higgs signal peaks with $M_A=125$ GeV and $\tb=30$ (leading to 
$M_h \sim 124$ GeV and $M_H \sim 134$ GeV) and backgrounds after 
detector resolution smearing; from Ref.~\cite{intense}.}
\label{Hdetect:icr}
\vspace*{-7mm}
\end{figure}

c) \underline{Intermediate coupling regime}: In the intermediate
coupling region,  that is for not too large $M_A$ values and
moderate $\tb \lsim 5$, the interesting decays $H \ra hh$, $A \ra
hZ$ and even $H/A \ra t\bar{t}$ [as well as the decays $H^\pm \to
Wh$] still have sizable branching fractions and can be searched
for.   In particular, the $gg \to H \to hh \to b\bar b \gamma
\gamma$ process (the $4b$ channel is more difficult as a result of
the large background) is observable for $\tb \lsim 3$ and $M_A \lsim
300$ GeV, and would allow to measure the trilinear $Hhh$ coupling.
These regions of parameter space may need to be reconsidered in the
light of the new Tevatron value for the top mass.

d) \underline{Intense--coupling regime}: In the intense--coupling
region, that is for $M_A \sim M_h^{\rm max}$ and $\tb \gg1$, the three
neutral Higgs bosons $\Phi=h,H,A$ have comparable masses and couple
strongly to isospin $-\frac{1}{2}$ fermions leading to dominant decays
into $b\bar b$ and $\tau\tau$ and large total decay widths
\cite{ICR,intense}. The three Higgs bosons can only be produced in the
channels $gg \to \Phi$ and $gg/q\bar q \to b\bar b + \Phi$ with $\Phi
\to b\bar b, \tau^+\tau^-$ as the interesting $\gamma \gamma, ZZ^*$
and $WW^*$ decays of the CP--even Higgses are suppressed. Because of
background and resolution problems, it is very difficult to resolve
between the three particles. A solution advocated in
Ref.~\cite{intense} (see also Ref.~\cite{ggmumu}), would be to search
in the channel $gg/q\bar q \to b\bar b + \Phi$ with the subsequent
decay $\Phi \to \mu^+ \mu^-$ which has a small BR, $\sim 3 \times
10^{-4}$, but for which the better muon resolution,  $\sim 1\%$, would
allow to disentangle between at least two Higgs particles.  The
backgrounds are much larger for the $gg \to\Phi \to \mu^+ \mu^-$
signals. The simultaneous discovery of the three Higgs particles is
very difficult and in many cases impossible, as exemplified in
Fig.~\ref{Hdetect:icr} where one observes only one single peak 
corresponding to $h$ and $A$ production. 

\begin{figure}[!h] 
\begin{center}
\mbox{
\includegraphics[width=7cm,height=5.5cm]{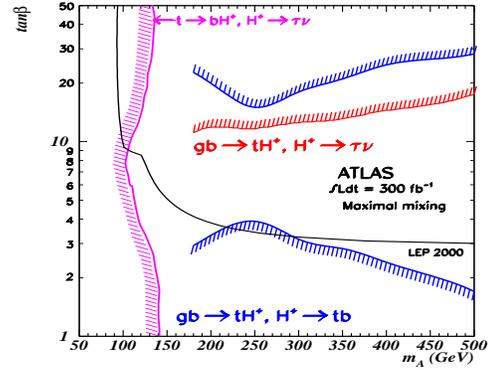} 
}
\end{center}
\vspace*{-9mm}
\caption{The coverage in the $M_A$--$\tb$ plane in the search for 
the charged Higgs boson at the LHC in  ATLAS simulations;
from Ref.~\cite{atlastdr}.}
\label{fig:H+LHC}
\vspace*{-2.mm}
\end{figure}

Finally, as mentioned previously, light $H^\pm$ particles with
masses below $M_{H^\pm} \sim m_t$ can be observed in the decays $t
\ra H^+b$ with $H^-\ra \tau \nu_\tau$; see Fig.~\ref{fig:H+LHC}.
Heavier $H^\pm$ states can be probed for large enough $\tb$, by
considering the properly combined $gb \to t H^-$ and $gg \ra t
\bar{b} H^-$ processes using the decay $H^-\ra \tau \nu_\tau$ and
taking advantage of the $\tau$ polarization to suppress the
backgrounds, and eventually the decay $H^- \to \bar{t}b$  which
however, seems more problematic as a result of the large QCD
background.  See Ref.~\cite{DP-H+} for more detailed discussions on
$H^\pm$ production and search strategies at the LHC.

\subsection{The SUSY regime}
\vspace*{1mm}

The previous discussion on MSSM Higgs production and detection at the
LHC might be significantly altered  if some sparticles are relatively
light. Some standard production processes can be affected, new
processes can occur and  additional channels involving SUSY final
states might drastically change the Higgs detection strategies. Let us
comment on some possibilities. 

The $Hgg$ and $hgg$ vertices in the MSSM are mediated not only by
heavy $t/b$ loops but also by loops involving squarks. If the top
and bottom squarks are relatively light, the cross section for the
dominant production mechanism of the lighter $h$ boson in the
decoupling regime, $gg \to h$, can be  significantly altered by
their contributions.  In addition, in the $h\to \gamma \gamma$ decay
which is one of the most promising detection channels, the same
$\tilde t, \tilde b$  loops together with chargino loops, will
affect the branching rate. The cross section times branching ratio
$\sigma( gg \ra h) \times {\rm BR}(h \ra \gamma \gamma)$ for the
lighter $h$ boson at the LHC can be thus very different from the SM,
even in the decoupling limit in which the $h$ boson is supposed to
be SM--like \cite{SUSYloops}. The effects can be drastic and could
lead to a strong suppression of  $\sigma(gg \ra h \ra \gamma
\gamma)$.

If one of the top squarks is light and its coupling to the $h$ boson
is  enhanced, an additional process might provide a new source for
Higgs  particles in the MSSM: associated production with
$\tilde{t}_1$ states,  $pp \ra gg/ q \bar{q} \ra \tilde{t}_1
\tilde{t}_1 h$ \cite{SUSYdirect}. This  is similar to the standard
$pp \to t\bar t h$ mechanism and in fact, for small masses and large
mixing of the $\tilde t_1$ the cross section can be comparable.
Final states with the heavier $H,A,H^\pm$ and/or other squark
species than $\tilde t_1$ are less favored by  phase space.

\begin{figure}[!h]
\vspace*{-.7cm}
\begin{center}
\epsfig{file=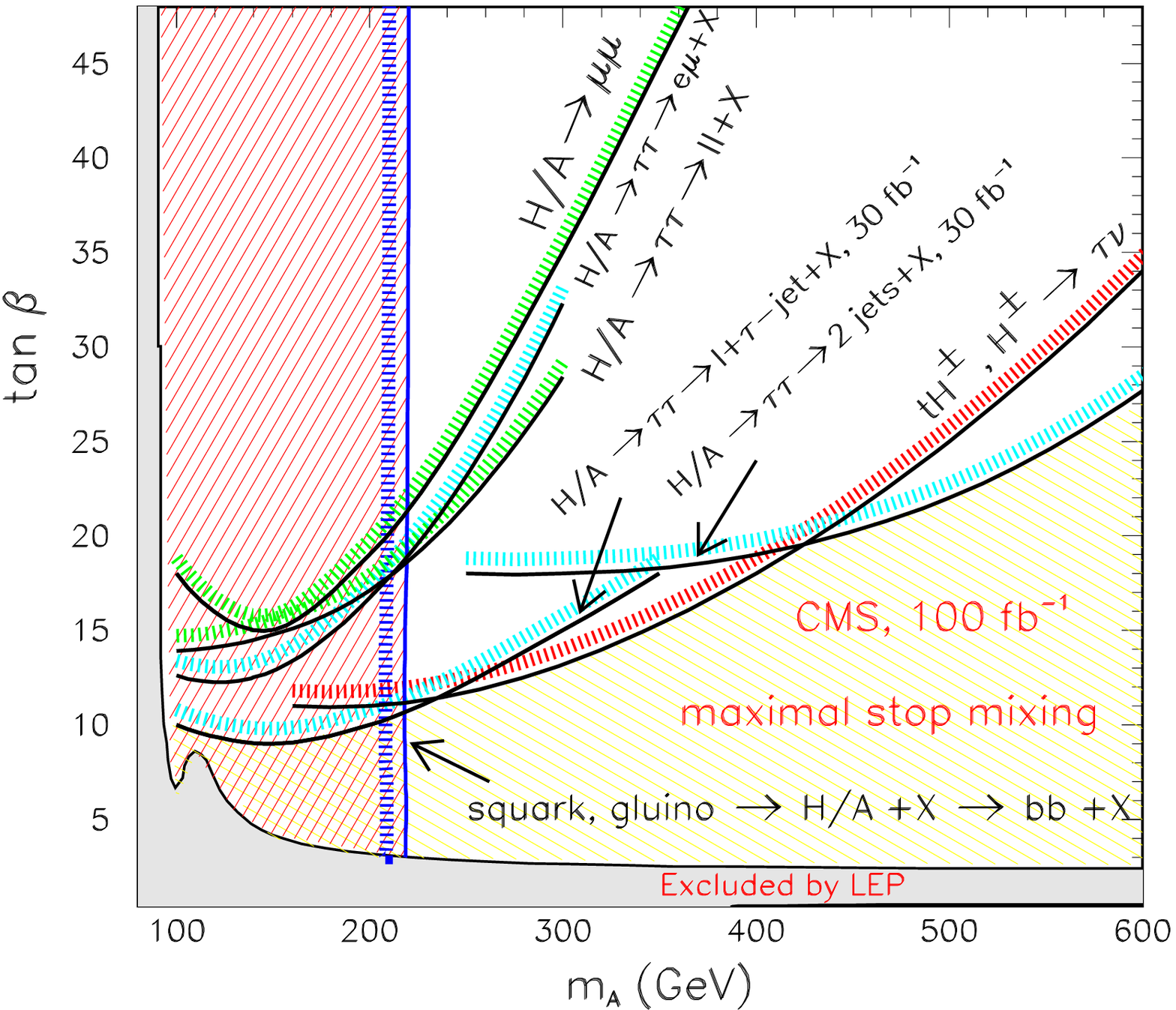,width=6cm} 
\end{center}
\vspace*{-1.cm}
\caption{Areas in the $[M_A, \tb]$ parameter space where the MSSM Higgs 
bosons can be discovered at the LHC with 100 fb$^{-1}$ data in 
sparticle cascades \cite{cascade}.} 
\label{SUSY:cascade}
\vspace*{-7.mm}
\end{figure}

Another possible source of  MSSM Higgs bosons would be from the 
cascade decays of strongly interacting sparticles, which have large
production rates at the LHC.  In particular, the lighter $h$ boson and
the heavier $A,H$ and $H^\pm$ particles with masses $\lsim 200$--300
GeV, can be produced from the decays of squarks and gluinos into the
heavier charginos/neutralinos, which then decay into the lighter ones
and Higgs bosons. This can occur either in ``little cascades",
$\chi_2^0, \chi_1^\pm \to \chi_1^0+\,$Higgs, or in ``big cascades"
$\chi_{3,4}^0, \chi_2^\pm \to \chi_{1,2}^0, \chi_1^\pm + \,$Higgs. The
rates for ino decays into Higgs bosons can be dominant while decays of
squarks/gluinos into the heavier inos are substantial.  Detailed
studies \cite{cascade} have shown  that these processes 
can be isolated in some areas of the SUSY parameter space and can
be complementary to the direct production ones; see
Fig.~\ref{SUSY:cascade}. In particular, one can probe the region $M_A
\!\sim\!150$ GeV and $\tb
\!\sim \!5$, where only  $h$ can be observed in 
standard searches.

\begin{figure}[!h]
\vspace*{-1.cm}
\begin{center}
\mbox{
\epsfig{file=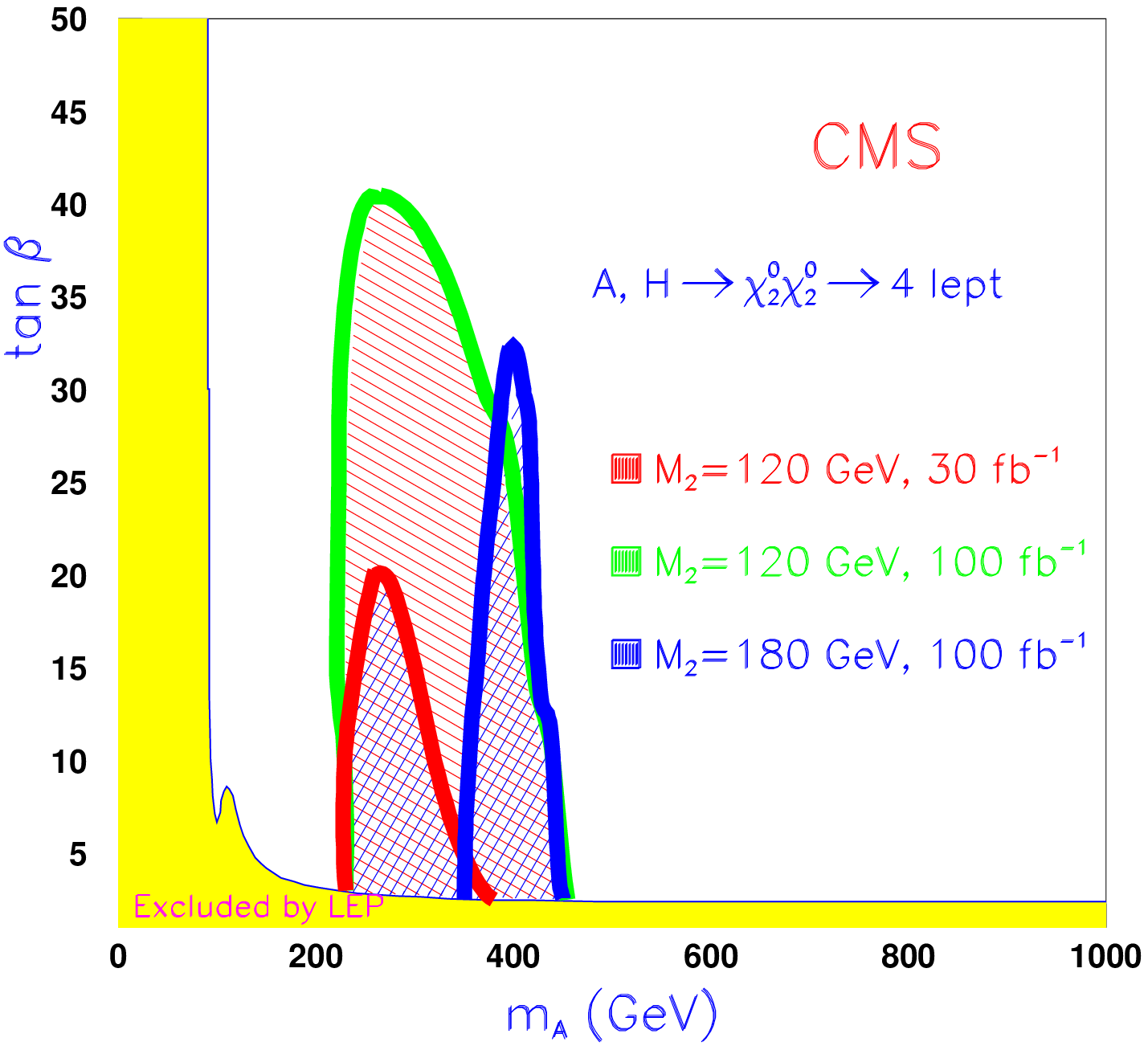,width=5.8cm} }
\end{center}
\vspace*{-11.mm}
\caption{Areas in the $[M_A, \tb]$ parameter space where the MSSM 
Higgs bosons can be discovered at the LHC with 100 fb$^{-1}$ data 
in $A/H \ra \chi^0_2 \chi^0_2\ra 4\ell^\pm +X$ decays for a 
given set of the MSSM parameters \cite{SUSY-Filip}.} 
\vspace*{-9.mm}
\label{SUSY:4leptons}
\end{figure}

The possibility of light charginos and neutralinos  allows to search
for the heavier $H,A$ and $H^\pm$ states in regions of the parameter 
space in which they are not accessible in the standard channels [this
is the  case e.g. for $M_A \sim 200$ GeV and moderate $\tb$ values].
There are situations in which  the signals for Higgs decays into
charginos and neutralinos are clean enough to be detected at the LHC. 
One of the possibilities is that the neutral $H/A$ bosons decay into
pairs of the second lightest neutralinos, $H/A \to \chi_2^0 \chi_2^0$,
with the subsequent decays of the latter into the LSP neutralinos and
leptons, $\chi_2^0 \to \tilde \ell^* \ell \to \chi_1^0 \ell \ell$ with
$\ell^\pm=e^\pm,\mu^\pm$, through the exchange of relatively light
sleptons. This leads to four charged leptons and missing energy in the
final state. If the  $H/A$ bosons are produced in the $gg$--fusion
processes, there will be little hadronic activity and the $4\ell^\pm$
final state is clean enough to be detected. Preliminary analyzes show
that  the decays can be isolated from the large (SUSY) background;
Fig.~\ref{SUSY:4leptons}.

\subsection{The CP violating MSSM}\smallskip

There are two ways by which CP violation in the MSSM affects the
Higgs production rates and detection at the LHC, through mixing 
in the Higgs states and/or modification of loop induced 
$gg H_i$ and $\gamma \gamma H_i$ couplings due to CP violation
in the squark couplings
\cite{cpnsh,mycpreview,CPHMSSM,HMSSMCPX,CPHmasses,dedes,moretti-Poulose,cpv-susyhiggs,ggr}.
Ref.~\cite{dedes}, for example, discusses 
the situation with no significant mixing between the $h,H$ and $A$
states and  large  effect on the $gg H_1$ coupling of CP violation in 
the squark-squark-Higgs vertex. 

\begin{figure}[!h]
\vspace*{-3mm}
\centerline{
\includegraphics[angle=90,width=4.1cm,clip=true]{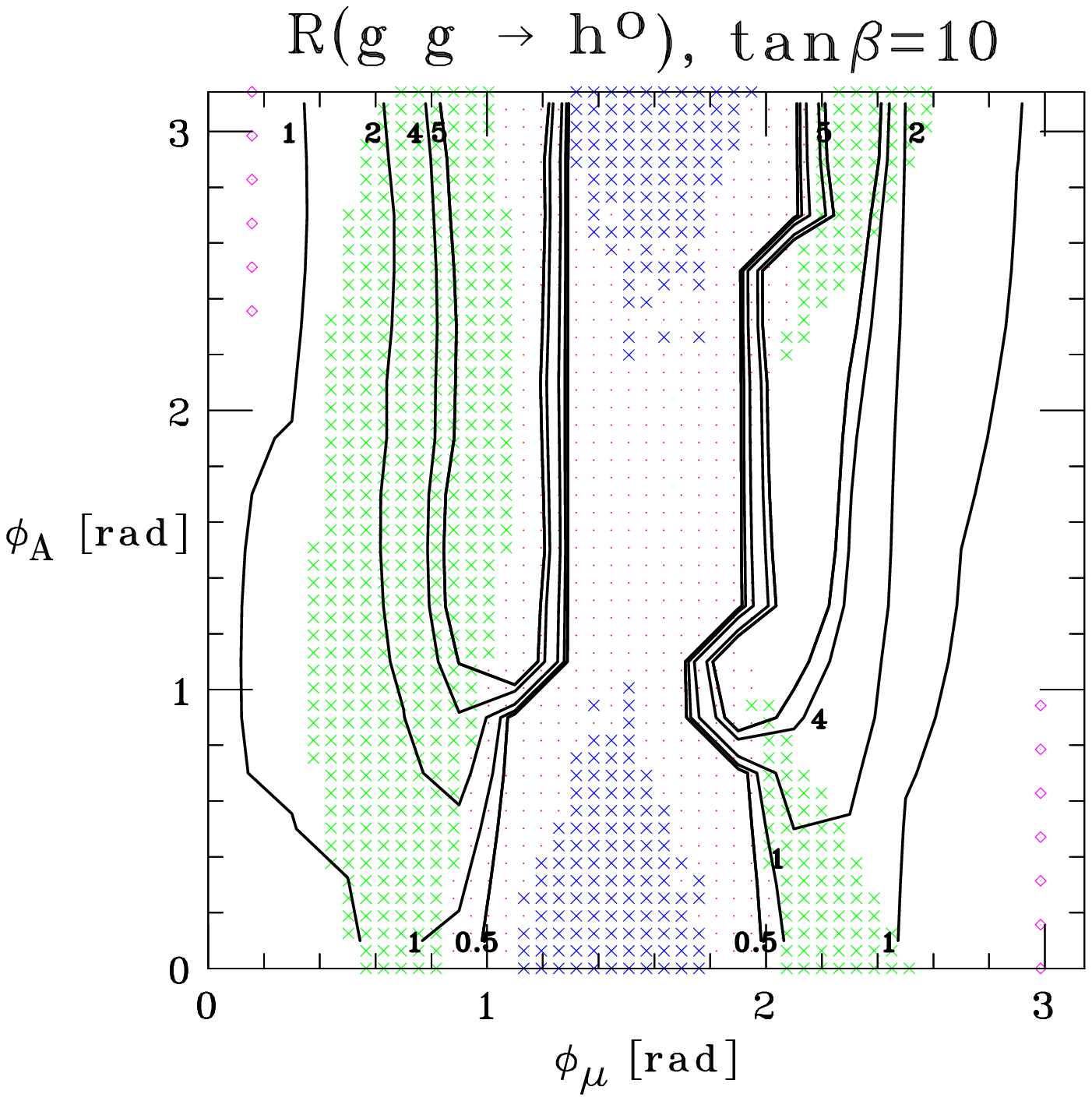}
\includegraphics[angle=90,width=4.1cm,clip=true]{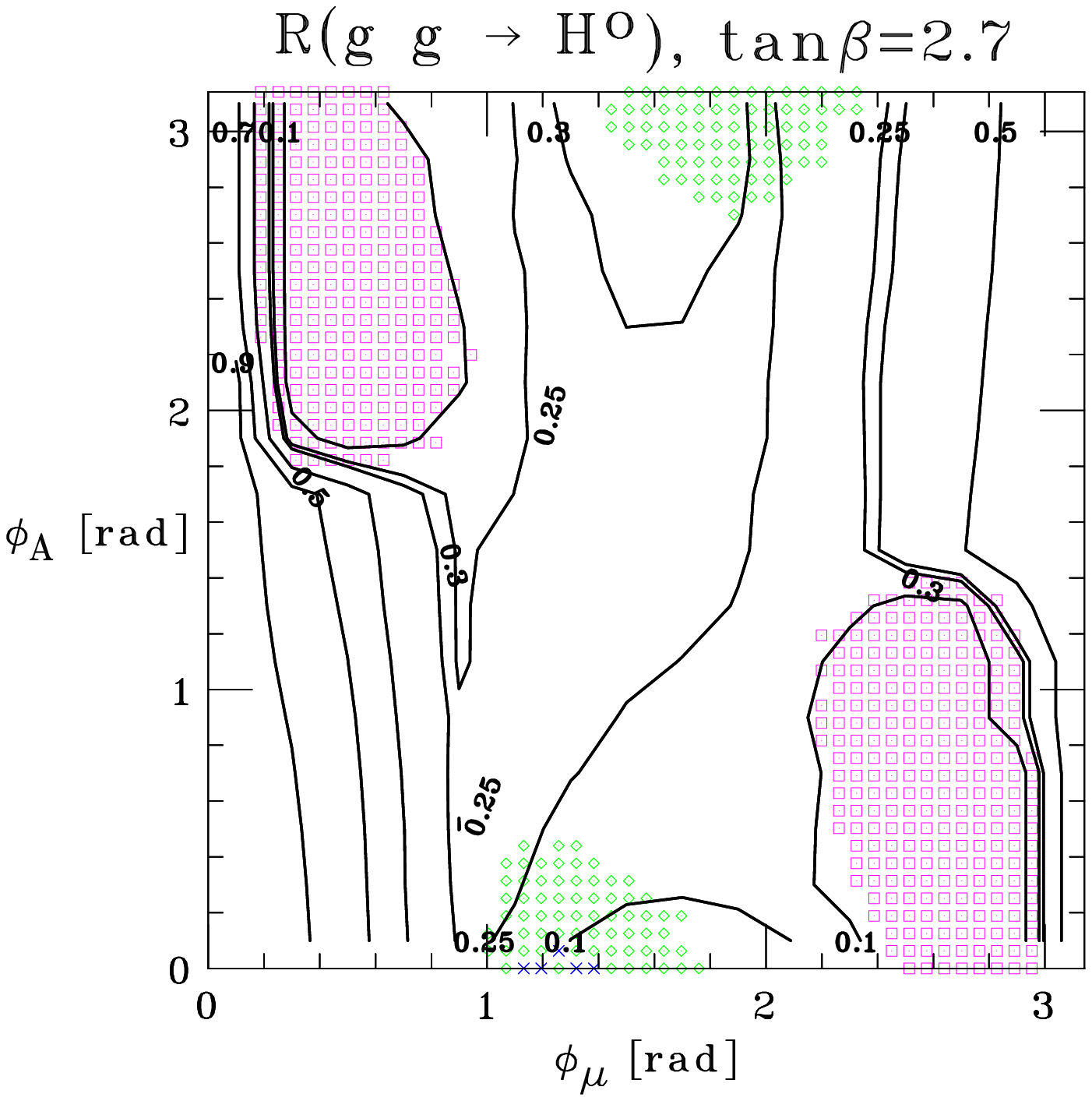}
}
\vspace*{-6mm}
\caption{Contours of ratio of Higgs production to that expected in the
CP conserving case, as a function of $\Phi_\mu$ and
$\Phi_A$\protect\cite{dedes}. The left panel is for $h$ and $\tan \beta =10$
and the right panel is for $H$ and for $\tan \beta = 2.7$. Also shown
are the regions disallowed by the EDM constraints.}
\label{dedeshiggs}
\end{figure}

In Fig.~\ref{dedeshiggs}, the contours of ratios of  $h,H$ production
rates in the CP violating MSSM to those without CP violation are
shown. This corresponds to the  case  where the CP violation  in the
MSSM induces CPV $\tilde q \tilde q h (H)$ couplings. As expected
from the sum rule we find that whereas the $h$ production rate
increase in the allowed region, the $H$ production rate decreases. As
can be seen from the Fig.~\ref{dedeshiggs} the effects can be
considerable.

A more interesting aspect of CP violation in the Higgs sector is the
vanishing  of the coupling of the lightest Higgs scalar $H_1$ to the
$Z$ pair mentioned  earlier, which in fact invalidates the lower
limit on the mass of the lightest neutral at the LHC.  Further, due
to the reduced $ggH_1$ and  $t \bar t H_1$ couplings in this case,
one may miss this state at the LHC too.   The region of the
parameter space where this happens, generically seems to correspond
to the case where all the three neutral scalars and  the charged
scalar are reasonably light. This region of Ref.~\cite{HMSSMCPX}
corresponds to $\tan \beta \sim 3.5-5,  M_{H^+}\sim 125-140 $ GeV,~
$ M_{H_1} \stackrel{<}{{}_\sim} 50 $ GeV and $\tan \beta \sim 2-3,
M_{H^+}\sim 105-130~{\rm GeV}, M_{H_1}  \stackrel{<}{{}_\sim} 40 $
GeV, for $\Phi_{CP} = 90^\circ$ and $60^\circ$ respectively. (The
details of the exact excluded region depend on the code used to
compute the spectrum~\cite{CPsuperH+FeynHiggs}).   An analysis
taking into account simulation of detector effects ~\cite{markus}
confirms that  there exists a  region in the $\tan \beta - M_{H^+}$
plane  corresponding to $M_{H_1} < 50$ GeV, $100 < M_{H_2} < 110$
GeV and  $130 < M_{H_3} < 180$ GeV~\cite{HMSSMCPX}, where LHC does
not seem to have  reach. 

\begin{figure}[!h]
\vspace*{-5mm}
\includegraphics*[width=7cm,height=5.8cm]{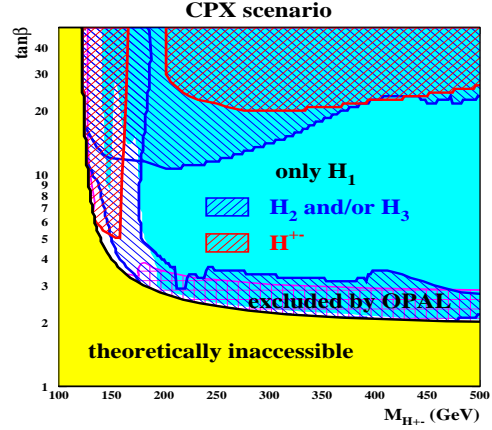}
\vspace*{-9mm}
\caption{Reach of LHC in $\tan \beta$--$M_{H^+}$  plane in the CPX
scenario~\protect~\cite{markus}} 
\label{Hdetect:CPXlhc} 
\vspace*{-7mm}
\end{figure}

\begin{figure}[!h]
\vspace*{-9mm} 
\includegraphics*[width=8cm]{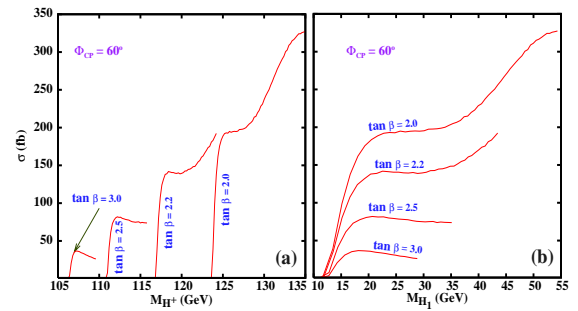}
\vspace*{-10mm}
\caption{Variation of the expected  cross-section with $M_{H^+}$  for
four values of $\tan\beta =2, 2.2, 2.5 $ and $3$. The CP-violating
phase $\Phi_{\rm CP}$ is $60^{\circ}$~\protect\cite{ggr}.} 
\label{Hdetect:cpvrate}
\vspace*{-7mm}
\end{figure}

This is shown in Fig.~\ref{Hdetect:CPXlhc}. In fact, the sum rules
that the $H_i$ couplings must satisfy, come to the rescue~\cite{ggr}
in recovering the lost Higgs signal. This is a very generic example
of how one can 'recover' the Higgs signal if the  model parameters
should contrive to make the usually investigated search channels
ineffective. $H_i VV$ and $H_i H^+ W$ couplings satisfy a sum rule
given by: $ g^2_{H_i VV} + |g_{H_i H^+ W}|^2 =  1$. Further, there
exists in the MSSM a correlation between the mass of the charged 
Higgs $M_{H^\pm}$ and that of the pseudo-scalar state. A suppressed
$H_1 VV$ coupling implies a light pseudo-scalar state,  which  in
turn implies a light charged Higgs, with $M_{H^+} < M_t$. Hence, a
light $H_1$ which might have been missed at LEP also corresponds to
an $H^\pm$ light enough to be produced in $t$ decay, which in turn
decays to $H_1 W^\pm$, followed by $H_1$ decay to $b \bar b$.  Due
to the large $H^\pm \rightarrow H_1 W^\pm$ branching ratio, the
expected event rate for this final state is quite healthy ($\sim {\cal O}
100$ fb), over the entire hole region; see Fig.~\ref{Hdetect:cpvrate}.

Thus one can look for the $H_1$ in final states containing $b W^+  b b
\bar b W^-$  in the $t \bar t$ sample. The huge background ($\sim 8.5
$ pb) coming from QCD production of $t \bar t \bar b b$,   can be
reduced to $\sim 0.5$ fb level and below, by demanding that one  of
the $b W$ combination reconstructs to $t$ mass and the $bbbW$ also to 
the $t$ mass~\cite{ggr,Houches-last}.  

\begin{figure}[!h]
\vspace*{-7mm}
\includegraphics[scale=0.3]{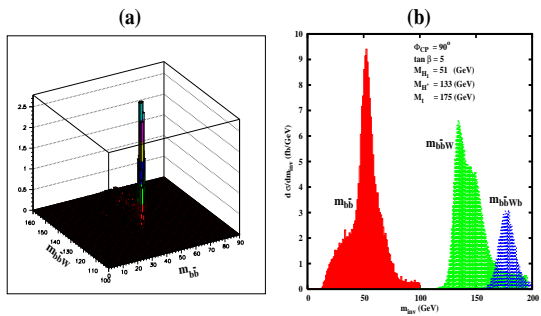}
\vspace*{-9mm}
\caption{Clustering of the $b\bar b, b\bar bW $ and $b\bar b b W$
invariant masses. $(a)$ three-dimensional plot for the correlation
between $m_{b\bar b} $ and $m_{b\bar b W}$ 
distribution. $(b)$ $m_{b\bar b}, m_{b\bar b W}$ and $m_{b\bar b W b}=
M_t$ distributions for $\Phi_{\rm CP} = 90^{\circ}$.
Appropriate $M_t, M_W$ mass window cuts have been applied  The other
MSSM parameters are $\tan\beta = 5, M_{H^+} = 133$ GeV, corresponding
to  $M_{H_1} = 51 $ 
GeV~\protect\cite{ggr}.} \label{Hdetect:cpvdistr} 
\vspace*{-2mm}
\end{figure}

Fig.~\ref{Hdetect:cpvdistr} shows the clustering of the $b\bar b,
b\bar bW $  and $b\bar b b W$ for the signal which can be used
effectively to handle the  background. These studies have thus opened
up a new topology in which one should be looking for the lightest
neutral Higgs in the decay of $H^\pm$  produced in $t$ decay.  A few
points  are worth noticing.  Due to the rather small value of $\tan
\beta$ the usual $\tau \nu_\tau$ decay mode for the $H^+$ is also not
available for the $H^+$ search in this case. Thus in this region of
the MSSM parameter space, the above process  provides a search
prospect not just for the light neutral state which might have been
missed at LEP, but also the light charged Higgs $H^+$ in this
parameter range (a similar  situation attains in NMSSM as
well~\cite{cpnshgodroy}). A  theorists analysis~\cite{adbiswarup}
indicates that it may be possible to look  at $\tilde t \bar {\tilde
t} H_1$ production, which will be higher than in the corresponding CP
conserving scenario due to lighter $H_1$, and have a signal for
parameter values corresponding to the hole. 


\subsection{Extensions of the MSSM}\smallskip

In the NMSSM, where a complex iso-scalar field is introduced, leading
to an additional pair of scalar and pseudoscalar Higgs particles,  the
axion--type or singlino  character of the pseudoscalar $A_1$ boson 
makes it preferentially light and decaying into $b$ quarks or $\tau$
leptons \cite{cpnsh,benchmark,Houches-last}. Therefore, in some areas of the
NMSSM parameter space, the lightest CP--even Higgs boson may
dominantly decay into a pair of light pseudoscalar $A_1$ bosons
generating four $b$ quarks or $\tau$ leptons  in the final state, $H_1
\to A_1 A_1 \to 4b, 2b2\tau, 4\tau$. In fact, it is also possible
that  $H_1$ is very light with small $VV$ couplings, while $H_2$ is
not too heavy and plays the role of the SM--like  Higgs particle; the
decays $H_2\to H_1 H_1$ can also be substantial and will give the same
signature as above.

\begin{figure}[!h]
\vspace*{-7mm}
\includegraphics*[width=7cm,height=5cm]{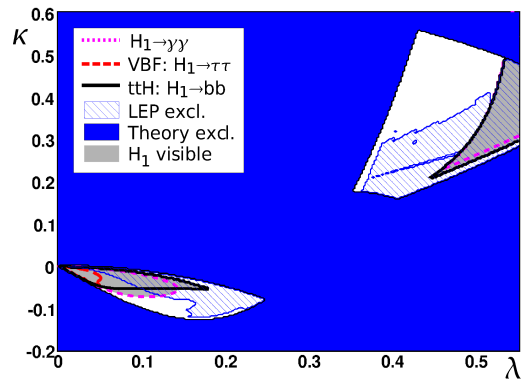}
\vspace*{-9mm}
\caption{Regions of the NMSSM parameter space $[\lambda,
\kappa$] in which a light pseudoscalar Higgs boson can be detected in an
ATLAS simulation \cite{Houches-last}.}
\label{Hdetect:nmssmlhc}
\vspace*{-7mm}
\end{figure}

This situation, similar to the CPX scenario discussed above, is very
challenging at the LHC. Indeed, all the production mechanisms of the
light $A_1$  or $H_1$ singlino--like state will have small cross
sections as both couplings to vector bosons and top quarks are tiny.
The SM--like Higgs $H_1$ or $H_2$ will have  reasonable production
rates but the dominant decay channels into $4b, 2\tau 2b$ and $4\tau$
will be swamped by the QCD background. Nevertheless, in the case  of
very light $A_1$ bosons with masses smaller than 10 GeV and, therefore
decaying almost exclusively into $\tau^+ \tau^-$ pairs, the  $H_1
\rightarrow A_1 A_1 \rightarrow 4\tau \rightarrow 4\mu + 4\nu_\mu
+4\nu_\tau$ final state with the $H_1$ boson dominantly produced in
vector boson fusion can be isolated in some cases. This is exemplified
in  Fig.~\ref{Hdetect:nmssmlhc} where the result of a simulation of
this process by members of the ATLAS collaboration is shown in  the
parameter space formed by the trilinear NMSSM couplings  $\lambda$ and
$\kappa$. While there are regions in which the final state can be
detected, there are other regions in which the light $H_1$ and $A_1$
states remain invisible even for the high luminosity which has been
assumed.

In the most general  SUSY model, with an arbitrary number of singlet
and doublet fields and an extended  matter content to allows for the
unification of the gauge couplings, a Higgs boson should have a mass
smaller than 200 GeV and significant couplings to gauge bosons and top
quarks; this particle can be thus searched for in the $gg$ and $VV$
fusion channels with the signature $WW \to \ell  \ell \nu \nu$ which
would be hard to miss.

Furthermore, in scenarios with  spontaneously broken R--parity,
besides invisible decays of the $h$ boson to be discussed later, 
decays of the pseudoscalar Higgs $A_i \to H_j Z \to Z$ and missing
energy could be detected if the cross sections for $A_i$ production
are large enough.

Other SUSY scenarios can also be probed at the LHC \cite{H:higheR,wess-rev}.  In
GUT theories which lead  to the presence of an extra neutral gauge boson at low
energies, the $Z'$  boson decays $Z' \to Zh$ which occur via $Z$--$Z'$ mixing
could have non--negligible rates and would lead to a detectable $\ell \ell b\bar
b$ signature; the $Z'$ production cross section would be large enough  for
$M_{Z'} \lsim 2$ TeV to compensate for the tiny mixing and hence, the small
$Z+$Higgs branching ratio.  If relatively light doubly charged Higgs bosons
exist, they can be produced in the Drell--Yan process $q\bar q \to H^{++}
H^{--}$  and, if their leptonic decays    $H^{--} \to \ell \ell$
are not too suppressed,  they  would lead to a spectacular 4--lepton final
state.

\subsection{Alternative scenarios and invisible Higgs}\smallskip

Various beyond the SM  physics options can in fact cause the Higgs
to have large branching ratio in ``invisible" final states: in the 
conventional MSSM, Higgs decays into LSP neutralinos $h \to \chi_1^0
\chi_1^0$, in the MSSM with R-parity violation decays into escaping
Majorons, $h \to JJ$,  mixing with graviscalars in extra dimensional
model are few of the reasons. In some cases this impacts the
branching ratio of the Higgs into the 'visible' final states such as
$b \bar b$ or $\gamma \gamma$ severely. The issue of how to search
for a Higgs  which dominantly decays into invisible decay products,
is therefore important from the point of view of recovering the lost
reach as well as for  measuring the invisible decay width. 

\begin{figure}[!h]
\vspace*{-.1mm}
\begin{center}
\includegraphics*[scale=0.4,angle=90]{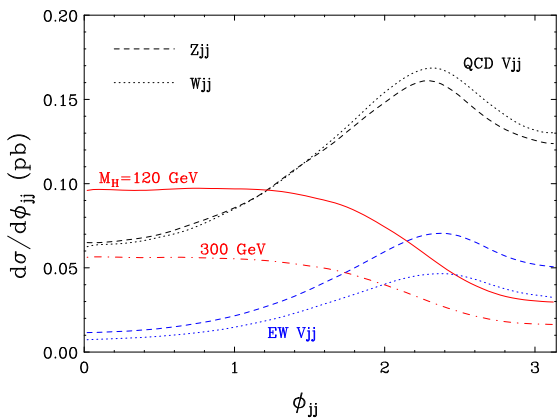} 
\end{center}
\vspace*{-12mm}
\caption{The azimuthal distribution between the jets for the signal for 
an invisibly decaying Higgs boson in the $WW$ fusion process
~\protect\cite{WWinvis} and background.}
\label{Hdetect:invisWWH}
\vspace*{-6mm}
\end{figure}

There have been many  parton level and  detector level studies on
this subject~\cite{gunioninvis,roy,WWinvis,guchait,han,invishou}.  The
most promising one is  the the production of the $h$ boson in the
$WW$ fusion  process, $q q  \rightarrow q q h$, which leads to   two
large rapidity jets with a rapidity gap ~\cite{WWinvis} along with
large missing momentum due to the invisible Higgs. 
Fig.~\ref{Hdetect:invisWWH} shows that the distribution in the
azimuthal  angle between the two jets, clearly distinguishes between
the dominant  $Z + 2$ jets background and the signal. With $100$
fb$^{-1}$ luminosity this method is shown to be sensitive for
invisible branching ratios as low as $5 \%$ ($12 \%$) for Higgs mass
$130$ ($400$) GeV.

\begin{figure}[!h]
\vspace*{-6mm}
\begin{center}
\includegraphics*[width=6cm,height=5cm]{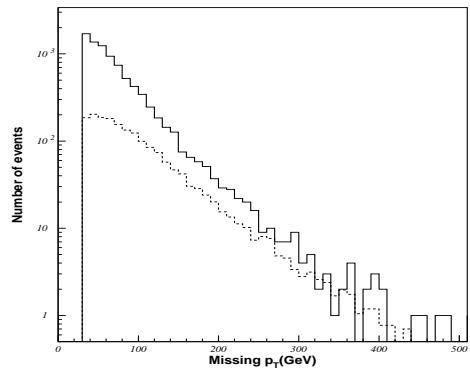}
\end{center}
\vspace*{-10mm}
\caption{Comparison of the $p_T$  distribution for the signal for 
the invisible Higgs in the $l^+ l^- + E_T^{\rm miss}$ 
channel~\protect\cite{guchait} and the irreducible $ZZ$ background.}
\label{Hdetect:invisZH}
\vspace*{-6mm}
\end{figure}

Alternatively, one can use the  production of a $h$ boson in
association with $Z$ boson followed by $Z$ decaying into a large
$p_T$ lepton pair with missing transverse energy
$E_T$~\cite{guchait}. Fig.~\ref{Hdetect:invisZH}  shows the  $p_T$
distribution for the signal (dashed histogram) and the dominant $ZZ$
background (solid histogram) in the case of the two-lepton
signature. This is a result of judicious cuts on different kinematic
variables exploiting the differences in the background and signal.
In fact, the cuts  significantly reduce the $Z + 2$ jets background
while affecting the signal only slightly. However, due to the small
production rate, this  process is useful only for large branching
ratios  ($40 \%$ or so) into the invisible channel and for Higgs
masses in the lower mass range.

Of course detection of an invisibly decaying Higgs boson will have
to be followed  by a study of attendant phenomenology predicted in
each of the corresponding models. Decays into LSP can not give rise
to a substantial  invisible branching ratio in the simplest mSUGRA
picture due to the current limits on chargino masses from LEP and
the attendant lower limits on the neutralino mass that exist in
these models. However, in the MSSM with non-universal  U(1) and
SU(2) gaugino masses $M_1$ and $M_2$,  it is possible to have
substantial invisible branching ratio corresponding to a light LSP
and still be consistent with the LEP results ~\cite{bbg4}. Further,
there still can exist regions of the parameter space where  $\sigma
(gg \rightarrow h \rightarrow \gamma \gamma)$, is  suppressed below
the value expected for the SM as previously discussed.

\begin{figure}[!h] 
\vspace*{-7mm}
\begin{center}
\includegraphics*[width=6.cm]{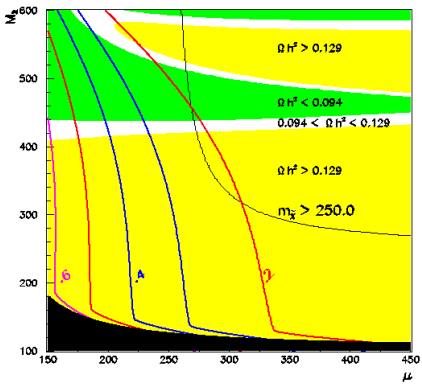} \end{center}
\vspace*{-15mm}
\caption{The invisible branching ratio for light Higgs in the
$M_2$--$\mu$  plane overlaid with regions allowed by relic density
constraints, for  non-universal gaugino masses $M_1/M_2 =
0.2$~\protect\cite{bbg4,Weiglein:2004hn}} \label{Hcosmo:bbgr} 
\vspace*{-7mm}
\end{figure}

In this case, it is the consistency with the cosmological relic
density of the LSP neutralino,  which requires that the small value of
the ratio $r= M_1/M_2$  be also accompanied by a light  slepton (which
in fact is preferred by the $(g-2)_\mu$ data), which constrains the
allowed region of this version of the MSSM. 
In this case,  the loss of the Higgs signal due to reduction in the
useful $\gamma \gamma$ and $b \bar b$ channels is compensated by
increased rate for production of $h$ in the decays of heavier
neutralinos and charginos caused by the fact that the LSP is a mixture
of gaugino and higgsino in this case. Fig.~\ref{Hcosmo:bbgr} shows this
connection between the Higgs sector  properties and the DM relic
density in the universe. The usual signal for the light Higgs in the
$\gamma \gamma$ final state is reduced here.

Invisible Higgs decays are also possible in non-SUSY models.  In
models with large extra dimensions \cite{LED}, the interaction  of
the Higgs field and the Ricci scalar curvature of the induced
four--dimensional metric also generates a mixing term with the
closest Kaluza--Klein graviscalar fields \cite{H-graviscalars}. This
mixing results in an effective Higgs decay width, $\Gamma(H \to 
{\rm graviscalar})$, which is invisible as the graviscalars are
weakly interacting and mainly reside in the extra dimension while
the Higgs is on the TeV brane. These invisible Higgs decays can be
largely dominating. In addition, there is the possibility of Higgs
decays into a pair of graviscalars, but the rates are smaller than
the ones from mixing.

Finally, let us comment on suppressed Higgs couplings in alternative
scenarios. As discussed previously, in Randall--Sundrum models
\cite{WED}, a scalar radion field is introduced to stabilize the
distance between the SM and the gravity brane. Carrying the same
quantum numbers, the Higgs and radion fields can mix and the
properties of the Higgs boson will be altered
\cite{Hewett:2002nk,Dominici:2002jv} and can lead to important  shifts
in the Higgs couplings which become apparent in the various decay
widths and production cross sections; see Fig.~\ref{fig:Hradion}. As
can be seen, while the shifts in the  $f \bar f/VV$ and $\gamma
\gamma$ widths are rather similar, the shift in the $H\to gg$ partial
decay  width is different; the width  can become close to zero for
some values of the mixing.  The impact of mixing in $f\bar f$ and $VV$
final states is in general smaller and the branching ratios will not
be significantly affected as these decays are dominant. This implies
that it will be imperative to perform a precise measurement of the
Higgs total decay width in order to probe the mixing  with radions. 

Another important consequence of radion mixing is the decays of the
Higgs boson into a pair of radions. Indeed, if the radion is
relatively light, the decays $H\to \phi \phi$ might be kinematically
accessible and, for some mixing values, the branching fractions might
be substantial. In some  mass range, e.g. $M_\phi \lsim 60$ GeV, the
radion will mainly decay into $b\bar b$ and $gg$ final states, while
the $\gamma \gamma$ branching ratio is very small. Observing these
final states will be rather difficult at the LHC. 

The suppression of the $Hgg$ loop induced coupling  can occur in  non
SUSY extensions of the SM as well. For instance, the ${\rm SU(2)_R}$
partner of the right--handed top quark in warped extra dimensional
models with an extended  left--right symmetric structure will also
contribute to the  $Hgg$ vertex and could interfere destructively with
the top quark contribution, leading to a much smaller coupling
\cite{H-RSHgg}. In the strongly interacting light Higgs scenario
proposed recently \cite{H-SILH}, the  Higgs couplings to gluons, as
well as  the couplings to fermions and gauge bosons, are also
suppressed.   The suppression of the $Hgg$ coupling  would lead to a
decrease  of the cross section for the dominant Higgs production
mechanism, $gg \to H$, and would make the Higgs search more
complicated at the LHC.

\section{Measurements of the Higgs properties}

It is clear from the discussion so far that after seeing the Higgs
signal at the LHC it will be essential to perform a measurement of
the Higgs properties, to be able to establish the exact nature of
EWSB and to achieve a more fundamental understanding of the issue. 
It is well known that a hadron  collider can afford only a limited
accuracy on measurements of most of the  Higgs properties and that
the next $e^+e^-$ linear collider  ILC will  indeed be needed for a
high precision measurement~\cite{DCR}.  Nonetheless, since LHC is
the current collider, it is important to address the Higgs
properties  question  when a large  luminosity, $\approx 300$
fb$^{-1}$, has been collected. We summarise some of the information
below.

\subsection{Mass, width and couplings of the SM Higgs}
\smallskip

The ease with which information can be obtained for the Higgs profile
clearly depends on the mass. The accuracy of the mass determination is
driven by the $\gamma \gamma$ mode for a light Higgs and by the $H
\rightarrow ZZ \rightarrow 4l$ mode for a heavier one and, in fact, is
expected to be accurate at one part in 1000. For  $M_H \gsim 500$
GeV,  the precision deteriorates rising to about a percent level
around $M_H \approx 800$ GeV, which is close to the theoretically 
expected upper limit,  due to  decreasing rates. 

\begin{figure}[!h]
\vspace*{-7mm}
\begin{center}
\includegraphics*[width=4.5cm,height=5cm]{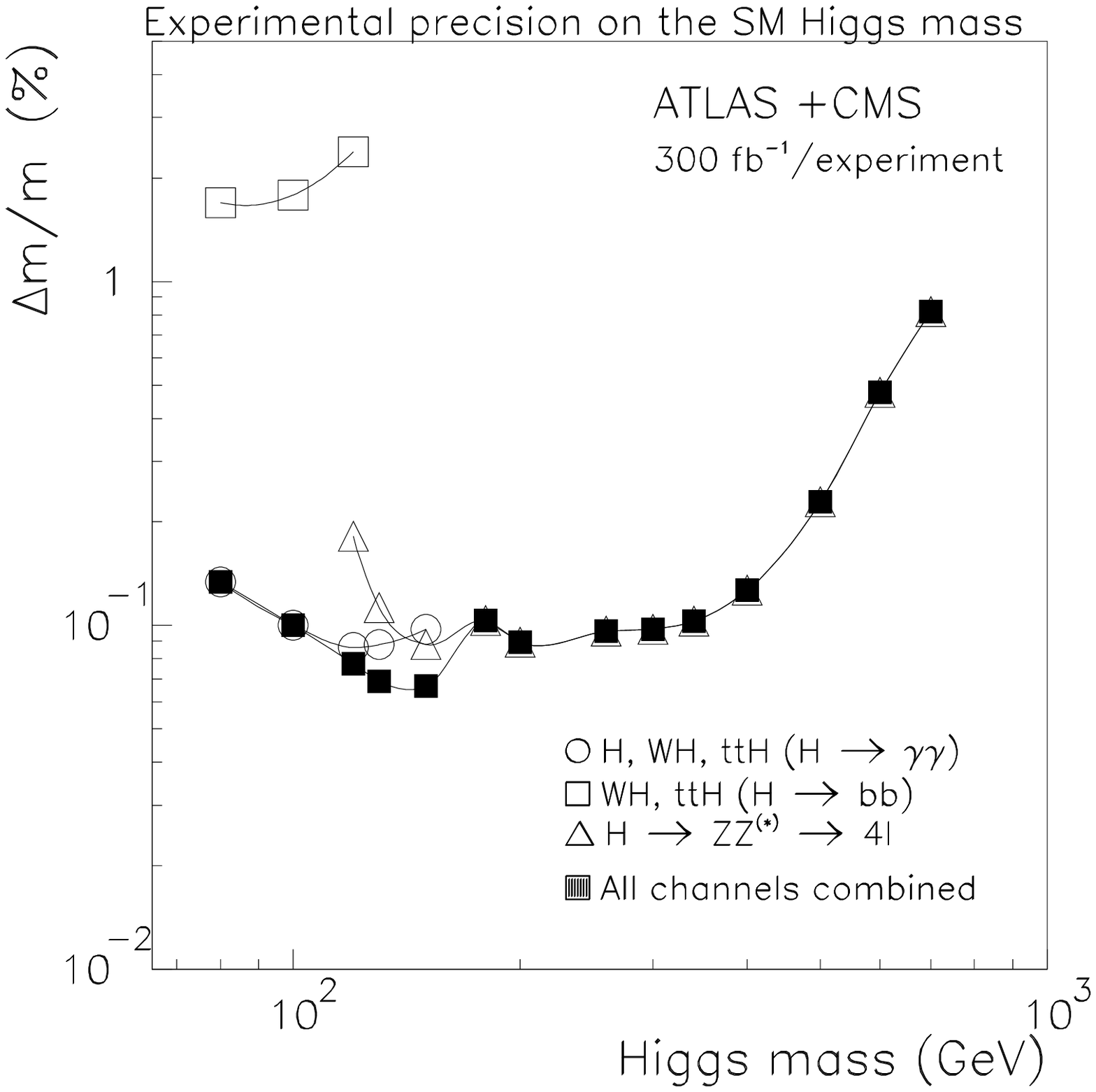}\hspace*{-5mm}
\includegraphics*[width=4.5cm,height=5cm]{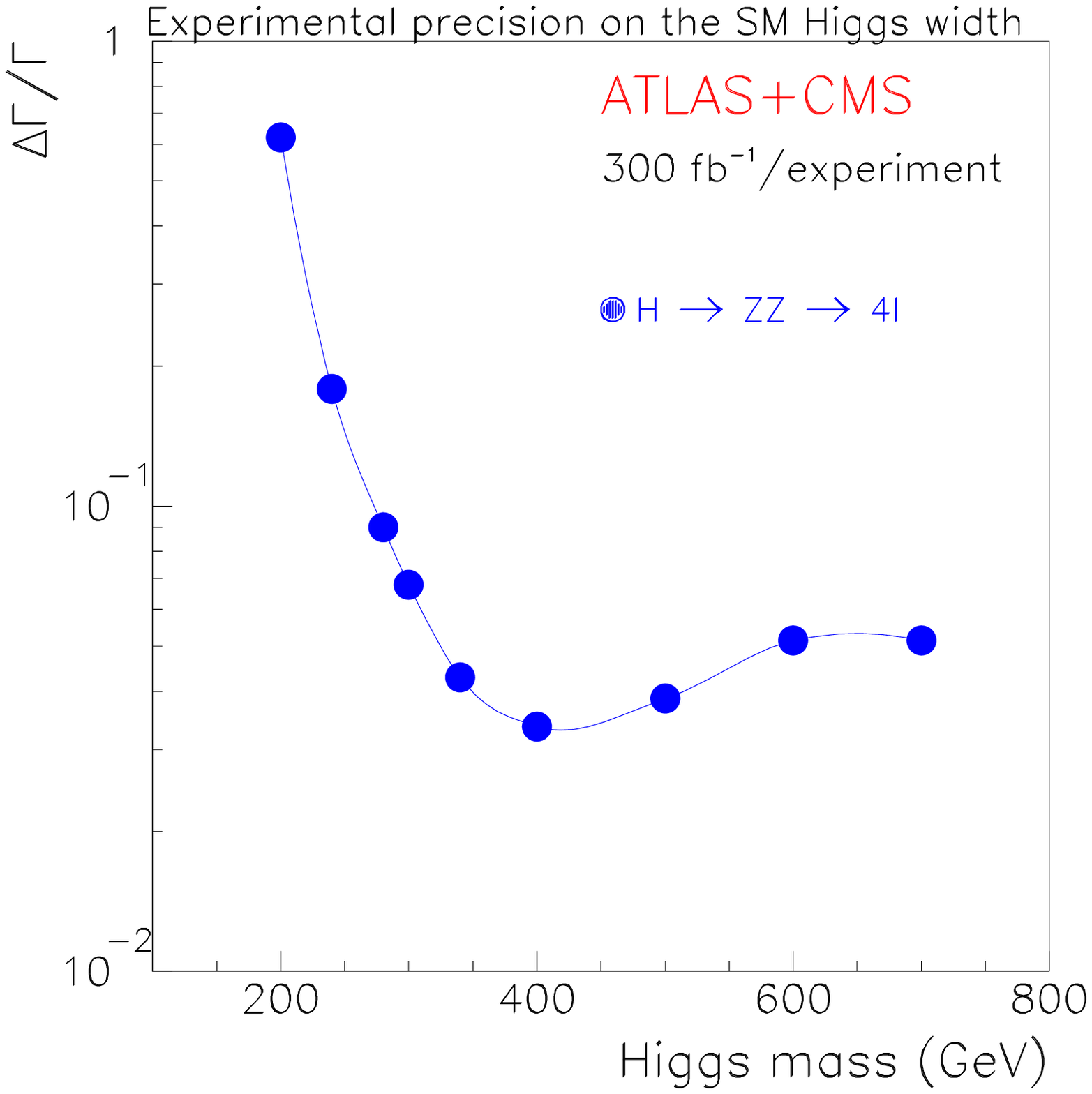}
\end{center}
\vspace*{-11mm}
\caption{Precision possible for the mass (left) and total width
(right) measurements for the SM Higgs for ${\cal L}=300$ fb$^{-1}$ 
combining ATLAS and CMS~\cite{atlastdr}.} 
\label{Hmass:preclhc}
\vspace*{-6mm}
\end{figure}

Using the same process, $H \to ZZ \to 4\ell^\pm$, the Higgs total
decay width can be measured for $M_H \gsim 200$ GeV when it is large
enough to be resolved experimentally. While the precision is rather
poor near this mass value,  it improves to reach the level of $\sim
5$\% around $M_H \sim 400$ GeV and the precision stays almost constant
up to masses of order $M_H\sim 700$ GeV \cite{atlastdr}.

\begin{figure}[!h]
\begin{center}
\includegraphics*[width=7cm,height=5cm]{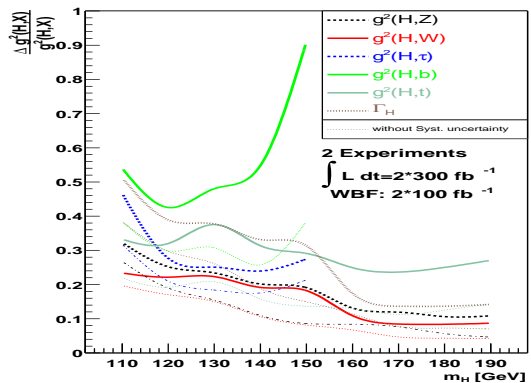}
\end{center}
\vspace*{-10mm}
\caption{Relative precision of fitted Higgs couplings-squared as a 
function of the Higgs mass for the $2 \times 300 + 2 \times 100$~fb
$^{-1}$ luminosity scenarios. It is assumed  that $g^2(H,V)<1.05 
\cdot g^2(H,V,SM)$ ($V=W,Z$) but one allows for new particles in the 
loops for $H\to\gamma\gamma$ and $gg\to H$ and for unobservable decay
modes~\protect\cite{Dieter1}.}
\label{Hcoup:preclhc}
\vspace*{-7mm}
\end{figure}

One would like to determine the couplings of the Higgs and test
their  proportionality to the masses of fermions/gauge bosons, which
is absolutely essential for checking the Higgs mechanism of EWSB.
Ratios of Higgs couplings squared can be determined by measuring
ratios of production cross sections times decay branching ratios 
and accuracies at the 10--50\% can be obtained in some cases
\cite{Dieter}. However, it has been shown in Ref.~\cite{Dieter1}
that with some theoretical assumptions, which are valid in general
for multi-Higgs doublet models, the extraction of absolute values of
the couplings rather than just ratios of the couplings, is possible
by performing a fit to the observed rates of Higgs production in
different channels. For Higgs masses below 200~GeV  they find
accuracies of order $10$--$40\%$ for the Higgs couplings after
several years of LHC running.  Fig.~\ref{Hcoup:preclhc} shows the
relative precision possible on fitted Higgs couplings-squared for
$2\!\times\! 300\! +\! 2\! \times\! 100$~fb$^{-1}$ as explained on
the figure. Thus at the LHC the various couplings can be determined
with a relative precision of at most $30 \% $. With just  $30$
fb$^{-1}$ data per experiment this is perhaps only good to $50$--$60
\% $ level. Ref.~\cite{Dieter1} also discusses how one can carry out
the program, for example, for the MSSM or for other beyond SM
models.

The trilinear Higgs boson self--coupling $\lambda_{HHH}$ is too
difficult to be measured at the LHC because of the smallness of the
$gg\to HH$ [and, {\it a  fortiori}, the $VV \to HH$ and $qq \to HHV$]
cross sections and the very large  backgrounds \cite{HHH-LHC,HHH-WW}.
A parton level analysis  has been recently performed in the channel
$gg\to HH \to (W^+W^-)(W^+W^-) \to (jj \ell \nu) (jj \ell \nu)$ and
$(jj \ell \nu) (\ell \ell \nu \nu)$ with same sign dileptons,
including all the relevant large backgrounds \cite{HHH-WW}. The
statistical significance of the signal is very small, even with an
extremely high luminosity,  and one can at most set rough limits on
the magnitude of the Higgs self-coupling.

Thus, for a very accurate and unambiguous determination of the Higgs
couplings, clearly an $e^+e^-$ Linear Collider~\cite{DCR} will be
required. 

\subsection{Measurements in the MSSM}
\smallskip

In the decoupling regime when $M_A \gg M_Z$, the measurements which
can be performed  for the SM Higgs boson with a mass $\!\lsim\! 140$
GeV will also be possible for the  $h$ boson.  Under some assumptions
and with 300 fb$^{-1}$ data,  coupling measurements would allow to
distinguish an MSSM from a SM Higgs particle at the $3\sigma$ level
for $A$ masses up to $M_A=$300--400 GeV \cite{Dieter}.  

The heavier Higgs particles $H,A$ and $H^\pm$ are accessible mainly in
the $gg\! \to\! b \bar b\!+\! H/A$ and $gb\!  \to\! H^\pm t$
production channels at large $\tb$, with the decays $H/A\! \to\!
\tau^+ \tau^-$ and $H^+\! \to\!  \tau^+ \nu$. The Higgs masses cannot
be determined with a very good accuracy as a result of the poor
resolution. However, for $M_A \lsim 300$ GeV and with high
luminosities, the $H/A$ masses can be measured with a reasonable
accuracy by considering the rare decays $H/A \to \mu^+ \mu^-$ 
\cite{intense,CMSTDR}. The discrimination between $H$ and $A$ is
though  difficult as the masses are close in general and the total
decay widths large \cite{intense}. 

\begin{figure}[h]
\vspace*{-7mm}
\begin{center}
\vskip 0.1 in
\includegraphics[width=70mm,height=50mm]{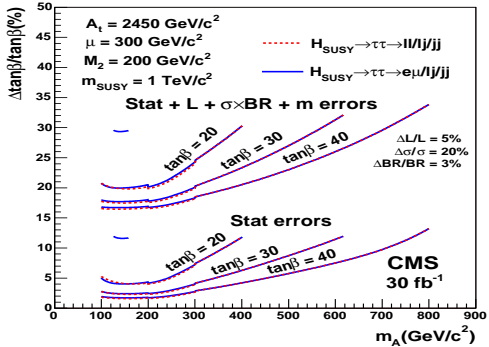}
\end{center}
\vspace*{-7mm}
\caption{The uncertainty in the measurement of $\tb$ in the 
channel $gg  \to H/A +b\bar b$ with the combined $H/A \to \tau \tau$ decays at 
CMS with 30 fb$^{-1}$ data. The three lower curves show the uncertainty when 
only statistical errors are taken into account, while the upper curves include
the uncertainties from the mass (a few \%) and luminosity (5\%) measurements 
and the theoretical uncertainty (23\%); from Ref.~\cite{Sasha-tb}.}
\vspace*{-7mm}
\label{Sasha}
\end{figure}

There is, however, one very important measurement which can be
performed in these channels. As the production cross sections above
are all proportional to $\tan^2\beta$ and, since the ratios of the
most important decays fractions are practically independent of $\tb$
for large enough values [when higher--order effects are ignored],
one has an almost direct access to this parameter.   A detailed
simulation shows that an  accuracy of $\Delta \tb/\tb \sim 30\%$ 
for $M_A\!\sim\! 400$ GeV and $\tb\!=\!20$ can be achieved with 30
fb$^{-1}$ data \cite{Sasha-tb}; Fig.~\ref{Sasha}. 

\subsection{Determination of the Higgs spin-parity}\smallskip

Apart from the mass, width and the couplings we also need to
determine the spin  of the Higgs and further establish that the Higgs
is a CP even  particle.  One can obtain information on these
properties by studying  various kinematical distributions such as the
invariant mass distribution  of the decay products and various
angular correlations among them, which depend on the spin of the
decaying object  crucially,  as well kinematical distribution of the
production process. A large amount of work  has been done on how to 
establish, at different colliders, that the Higgs boson is indeed  
${\rm J^{PC} = 0^{++}}$
state \cite{cpnsh,Godbole:2004xe}.  Most of the  analyses/suggestions for
the LHC emanate by translating the strategies devised in the case of the ILC.

\begin{figure}[!h]
\vspace*{-3mm}
\begin{center}
\includegraphics*[width=7cm,height=5cm]{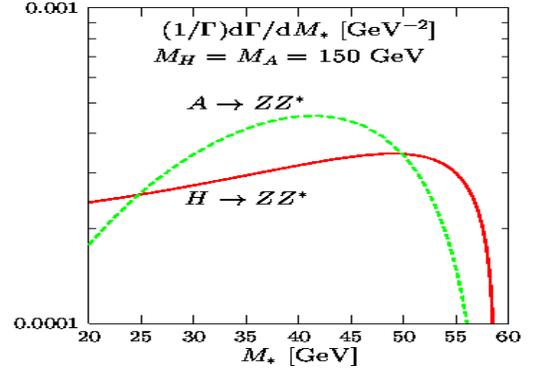}
\end{center}
\vspace*{-11mm}
\caption{Dependence on the CP quantum number of the Higgs for  the 
threshold behaviour of the distribution in $M_{Z^*}$ for the 
$H \rightarrow ZZ^*$ decay~\cite{Bargeretal}.}
\label{Hcp:lhccp}
\vspace*{-7mm}
\end{figure}

\begin{figure}[!h]
\vspace*{-7mm}
\begin{center}
\includegraphics*[width=7cm,height=5cm]{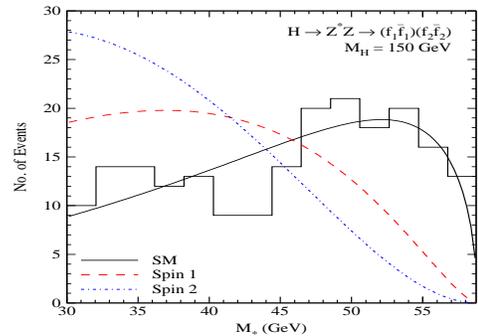}
\end{center}
\vspace*{-11mm}
\caption{Spin determination of the Higgs boson via the threshold
behaviour of the distribution in $M_{Z^*}$ for the $H \rightarrow
ZZ^*$ decay~\cite{Choi:2002jk}.}
\label{Hcp:lhcspin}
\vspace*{-7mm}
\end{figure}

One example is to study the threshold behaviour of the $M_{Z^*}$
spectrum in the $H \rightarrow Z Z^{(*)}$ decay for $M_H \lsim 2M_Z$.
Since the relative fraction of the longitudinally to transversely
polarised $Z$ varies with $M_{Z^*}$, this distribution is sensitive to both the
spin and the CP property of the Higgs. This is seen in
Figs.~\ref{Hcp:lhccp} and \ref{Hcp:lhcspin} where the behaviors
for a CP-even and CP-odd states  and for different spins are shown 
respectively.

Another very useful diagnostic of the CP nature of the Higgs boson is
the azimuthal distribution between the decay planes of the
two lepton pairs arising from the $Z, Z^{(*)}$ bosons coming from the
Higgs  decay
\cite{cpnsh,Bargeretal,Choi:2002jk,Buszello:2002uu,Allanach:2006yt,gmm}. 
Alternatively, one can study the distribution in the azimuthal angle
between the two jets produced in  association with the Higgs produced
in  vector boson fusion
\cite{Plehn:2001nj,Zhang:2003it,Buszello:2006hf} or in gluon fusion
in Higgs plus jet events~\cite{DelDuca:2001ad,Hankele:2006ja}.

\begin{figure}[!h]
\begin{center}
\includegraphics*[width=7cm,height=5cm]{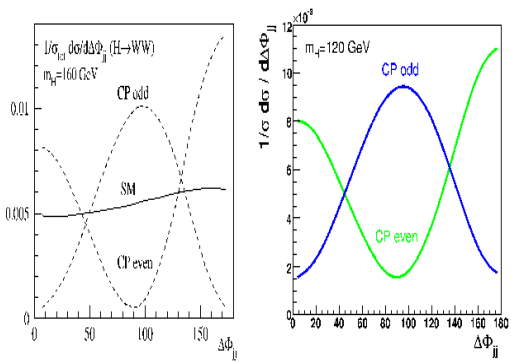}
\end{center}
\vspace*{-11mm}
\caption{Azimuthal angle distribution for the two jets produced in 
association with the Higgs boson, for CP-even and odd cases. Left shows 
the vector boson fusion case, for $M_H = 160$ GeV and  right, the 
gluon fusion for a mass $M_H = 120$ GeV~\protect\cite{Hankele:2006ja}.}
\label{Hcp:lhcjtcp}
\vspace*{-6mm}
\end{figure}

\begin{figure}[!h]
\begin{center}
\includegraphics*[width=8cm,height=5cm]{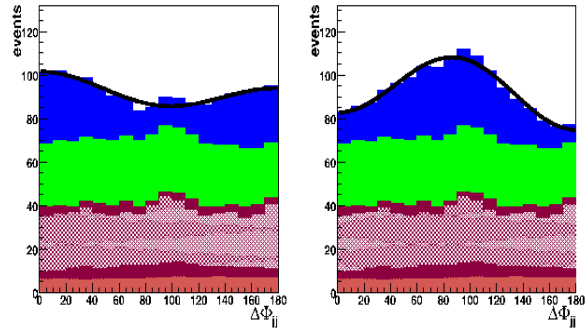}
\end{center}
\vspace*{-11mm}
\caption{Azimuthal angle distribution for the two jets produced in 
association with a Higgs, for the CP even (left) and CP odd 
(right) cases, after  selection cuts~\cite{Hankele:2006ja}, for 
$M_H = 160$ GeV. Shown are the gluon signal and the other 
backgrounds from top to bottom.}
\label{Hcp:lhcjtcpsim}
\end{figure}

Figures~\ref{Hcp:lhcjtcp} and \ref{Hcp:lhcjtcpsim}, show the
azimuthal angle distribution for the two jets produced in association
with the Higgs, for the CP--even and CP--odd cases, for the vector
boson and gluon  fusion, and  the gluon signal along with
vector boson contribution and all the other backgrounds,
respectively. One can see that  with a high luminosity of $300$
fb$^{-1}$, it should be possible to use these processes quite
effectively. Recall, however, that any determination of the CP
property using a process  which involves  the coupling of the spin 0
particle to a pair of gauge bosons, is ambiguous as only the CP even
part of the  coupling is projected out.

Couplings of a Higgs with heavy fermions offer therefore the best
option. $t\bar t$ final states produced in the decay of an
inclusively produced Higgs can be used to  obtain information on the
CP nature of the $t\bar t H$ coupling through spin-spin
correlations~\cite{Bernreuther:1997gs,Khater:2003wq}. Using optimal
observable analyses, the associated  $Ht\bar t$ production   allows a
determination of the CP-even and CP-odd part of the $t \bar t$
couplings with the Higgs boson separately~\cite{Gunion:1996xu},
though it requires  high luminosity.   The use of $\tau$ polarisation
in resonant $\tau^+ \tau^-$ production at the LHC has also been
recently investigated~\cite{HCPR1}. A novel 
approach~\cite{Khoze:2001xm,Ellis:2005fp}, is to use double-diffractive 
processes with large rapidity gaps where only scalar Higgs production is 
selected. 

\begin{figure}[!h]
\vspace*{-7.mm}
\includegraphics*[width=4.7cm,height =6.3cm,angle=90]{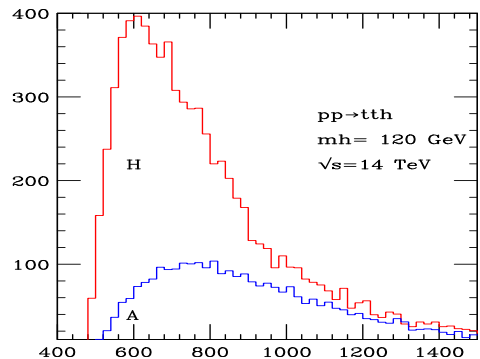}
\vspace*{-7mm}
\caption{Distribution in the $t \bar t\,$Higgs invariant mass 
for $pp \rightarrow t \bar t\,$Higgs for scalar and a pseudoscalar 
bosons $H$ and $A$ of 120 GeV at the LHC \cite{Fabio-ggttH}.}
\label{Hcsec:tthcp}
\vspace*{-9mm}
\end{figure}

In fact, recently, it was observed that the threshold rise of
$\sigma( e^+ e^- \rightarrow t \bar t)+$ Higgs at the ILC offers a
very clear and unambiguous determination of the CP nature of the $t
\bar t\,$Higgs coupling~\cite{Bhupal Dev:2007is}. The very different
rise of the cross-section with the $t \bar t\,$Higgs invariant mass
away from the threshold,  $2 M_t + M_H$, can be completely understood
in terms of  angular momentum  and parity conservation. 
Interestingly, the same is found to also hold for $gg \rightarrow t
\bar t+$Higgs production as well; see Fig.~\ref{Hcsec:tthcp} \cite{Fabio-ggttH}. 

Most of the suggested measurements should be able to verify  the
CP nature of a Higgs boson  when the full luminosity of
$300\,$fb$^{-1}$ is collected at the LHC or even before, provided the
Higgs boson is a CP eigenstate. However, a measurement of the CP
mixing is much more difficult, and a combination of several different
observables will be essential.\smallskip

 The subject of probing CP mixing reduces more generally to the 
probing of the anomalous $VV H$ and $t \bar t H$ couplings, the only
two cases where such study can even be attempted at the LHC. Since
CP--even and CP--odd Higgs bosons couple to the $t \bar t$ pair 
democratically where as the  coupling to a $VV$ pair is suppressed for
the CP--odd case, the most unambiguous in this context will be 
the  $t \bar t H$ process~\cite{Gunion:1996xu}. However, as already
mentioned a CMS study shows that, at present, it is  not clear whether
it would be possible to detect the $t \bar t H$ signal above the $t
\bar t b \bar b$ background. Hence, $VVH$ is the only relevant case. In
principle, the same studies which are used to determine the CP-even
or CP-odd character of the Higgs boson mentioned above, can be used.

\begin{figure}[!h]
\begin{center}
\includegraphics*[width=7cm]{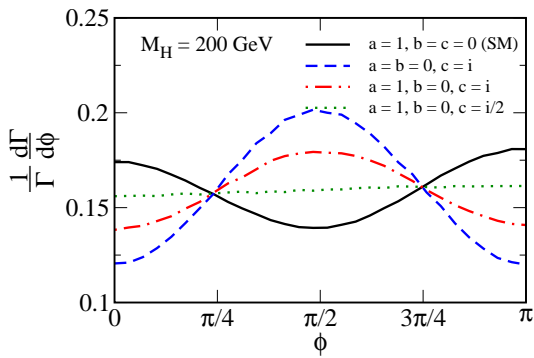}
\end{center}
\vspace*{-11mm}
\caption{The normalized differential width for \mbox{$H \rightarrow
Z^{(*)} Z \rightarrow (f_1\bar{f}_1) (f_2\bar{f}_2)$} with respect to
the azimuthal angle $\phi$. The solid (black) curve shows the
SM case ($a=1$, $b=c=0$) while the dashed (blue) curve is a
pure CP-odd state ($a=b=0$, $c=i$). The dot-dashed (red) curve and the
dotted (green) curve are for states with CP violating couplings $a=1$,
$b=0$ with $c=i$ and $c=i/2$, respectively~\cite{gmm}.}
\label{Hcp:mixphilhc}
\vspace*{-7mm}
\end{figure}

As an example, we show in Fig.~\ref{Hcp:mixphilhc}
the distribtuion in azimuthal angle $\phi$,
for $M_H = 200$ GeV,
where $H$ corresponds to a Higgs which may have indeterminate CP
assignments.  It should be kept in mind though, that this method
cannot be applied for very large Higgs masses where this
dependence is washed out. One must also beware of degenerate
Higgs bosons of opposite CP; since the decay products are the same,
they will both contribute to the rate and must be summed coherently,
possibly mimicking the effect seen above. Also, in the  context of
LHC with the QCD environment and modifications to shapes of
distributions, it is useful to construct specific observables which
may be directly proportional to the anomalous part of the coupling,
as was done in the $e^+e^-$ case recently~\cite{vvhanom}.

Parameterising the anomalous vertex by
\begin{eqnarray}
 V_{HZZ}^{\mu \nu} \, =\,
\frac{ig m_Z}{\cos\theta_W} \left[ \,a\, g_{\mu\nu}
+  b \,\frac{p_\mu p_\nu}{m_Z^2}
+  c \,\epsilon_{\mu\nu\alpha\beta} \, \frac{ p^\alpha k^\beta}{m_Z^2}
\, \right], \nonumber
\label{param}
\end{eqnarray}
where  $p=q_1+q_2$ and $k=q_1-q_2$, $\theta_W$ denotes the
weak-mixing angle and $\epsilon_{\mu \nu\alpha\beta}$ is the totally
antisymmetric tensor with $\epsilon_{0123}=1$, one can develop a
strategy to probe different parts of the anomalous couplings $a,b$
and $c$ directly. The general strategy is to construct different
observables out of the available 4-momenta such that they have
specific CP and $\rm \tilde T$ transformation properties, using
partially integrated cross-sections, where $\rm \tilde T$ denotes
naive time reversal. Then, the expectation value of the sign of this
observable, which will correspond to partially integrated 
cross-sections will be directly proportional to  the particular
anomalous coupling (or product of these)  which have  the same
transformation property. In the reasonable approximation of small
anomalous couplings, these observables will then directly probe
different anomalous couplings

\begin{figure}[!h]
\vspace*{-7mm}
\begin{center}
\includegraphics*[width=7cm,height=4.7cm]{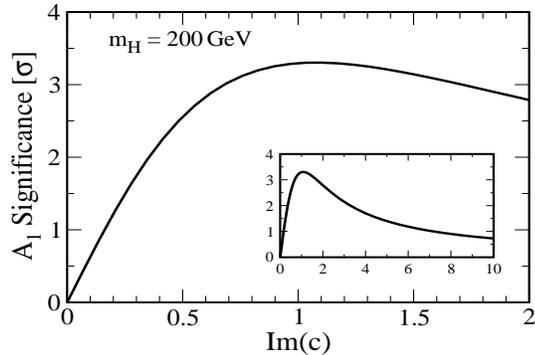}
\end{center}
\vspace*{-11mm}
\caption{The significances corresponding to the asymmetry ${\cal A}_1$
as a function of $\Im m(c)$, for a Higgs boson of mass
200~GeV.  We chose the CP-even
coupling coefficient $a=1$ and $b=0$. The inserts show
the same quantities for a larger range of $\Im m(c)$~\cite{gmm}.}
\label{Hcp:mixcpa1}
\vspace*{-7mm}
\end{figure}

One example of such an observable is the cosine of the angle
$\theta_1$ made by the decay  lepton with the $Z$ direction in the
rest frame of the Higgs boson.  One  can write
$$
O_1 \equiv \cos \theta_1  = \frac{(\vec{p}_{\bar f_1} - \vec{p}_{f_1})
\cdot (\vec{p}_{\bar f_2} + \vec{p}_{f_2})}{|\vec{p}_{\bar f_1} -
\vec{p}_{f_1}| |\vec{p}_{\bar f_2} + \vec{p}_{f_2}|}
$$
for the decay  $H \rightarrow Z Z^{(*)} \rightarrow f_1 \bar f_1
f_2 \bar f_2$. The expectation value of the sign is
$$
{\cal A}_1 = \frac{\Gamma (\cos\theta_1 > 0)-\Gamma (\cos\theta_1<0)}
{\Gamma (\cos\theta_1 > 0)+\Gamma (\cos\theta_1<0).}
$$
This is $\propto \Im m (c)$ and is thus a direct probe of nonzero
value for it and hence of CP violation. For example, for $M_H = 200$
GeV, values of $A_1$ of about $8 \%$ are possible.
Fig.~\ref{Hcp:mixcpa1} shows the sensitivity of such a measurement
for ATLAS with an integrated luminosity of $300$ fb$^{-1}$ for a
scalar of mass 200 GeV.
One can, in fact,  systematically construct observables, using this
strategy, to  probe the different parts of the anomalous
couplings separately. Thus, in principle, with high luminosities it
will be possible to map the anomalous $HZZ$ couplings at the LHC at
the level of $40$--$50 \% $. Of course, this precision is no comparison 
to what will be achievable at the ILC, as can be seen, for example,
from recent discussions in Refs.~\cite{vvhanom,Dutta:2008bh,DCR}.

In short, all the discussions above indicate that while LHC  with
$300$ fb$^{-1}$ data per experiment can perform measurements of
different Higgs properties, it is really to the ILC~\cite{DCR} that we
have to look to for precision information. 

\section{Conclusion}

The LHC will tell!

\subsection*{Acknowledgments:}  We acknowledge support from the Indo-French
IFCPAR under  project no: 3004-2 ``SUSY, Higgs and CP at colliders and in
astrophysics".  A.D. acknowledges support from  the Alexander von--Humbold
Foundation (Bonn, Germany).  R.G. would like to acknowledge  support
from the Department of Science and Technology, India
under the J.C.Bose Fellowship.

\vspace{-.3cm}


\end{document}